\documentclass{aa}
\usepackage[pdftex]{graphicx}
\usepackage{epsfig}
\usepackage{amsmath}
\usepackage{wasysym}

\usepackage{marvosym}
\usepackage{txfonts}
\pdfoutput=1
\usepackage[pdftex]{color,xcolor}
\usepackage[pdftex]{graphicx}
\usepackage{hyperref}
\hypersetup{pdfauthor=Christoph Mordasini}
\hypersetup{backref=true, pagebackref=true, hyperindex=true, breaklinks=true,colorlinks=true,urlcolor=blue, linkcolor=blue,  citecolor=blue,pagecolor=red, bookmarks=true, bookmarksopen=true}
\usepackage{epstopdf}

\def\mearth{M_\oplus}
\def\rearth{R_\oplus}

\def\msun{M_\odot}

\def\mcore{M_{\rm core}}

\def\f1{f_{\rm I}}
\def\mj{M_{\textrm{\tiny \jupiter }}}
\newcommand{\lj}{L_{\textrm{\tiny \jupiter}}}
\newcommand{\rj}{R_{\textrm{\tiny \jupiter}}}

\def\mstar{M_*}

\def\rcore{R_{\rm core}}
\def\mdotcore{\dot{M}_{\rm core}}

\def\beq{\begin{equation}}
\def\eeq{\end{equation}}

\def\fopa{f_{\rm opa}}
\def\t2{\tau_{\rm II}}

\def\sigmas0{\Sigma_{\rm s,0}}

\newcommand{\rcapt}{R_{\rm capt}}
\newcommand{\mz}{M_{\rm Z}}
\newcommand{\mxy}{M_{\rm XY}}

\newcommand{\lsun}{L_{\odot}}

\newcommand{\mdotz}{\dot{M}_{\rm Z}}
\newcommand{\mdotxy}{\dot{M}_{\rm XY}}

\def\araa{ARA\&A}             
\def\apj{ApJ}                 
\def\apjl{ApJ}                
\def\apjs{ApJS}               
\def\aap{A\&A}                
\def\mnras{MNRAS}             
\def\pasp{PASP}               
\def\ssr{Space~Sci.Rev.}      

\def\({\left(}
\def\){\right)}
\def\<{\left<}
\def\>{\right>}

\begin{document}

\title{Characterization of exoplanets from their formation I:\\ Models of combined planet formation and evolution}

\author{C. Mordasini\inst{1} \and  Y. Alibert\inst{2}   \and H. Klahr\inst{1}  \and T. Henning\inst{1} }

\institute{Max-Planck-Institut f\"ur Astronomie, K\"onigstuhl 17, D-69117 Heidelberg, Germany \and
Center for space and habitability, Physikalisches Institut, University of Bern, Sidlerstrasse 5, CH-3012 Bern, Switzerland}

\offprints{Christoph MORDASINI, \email{mordasini@mpia.de}}

\date{Received 15.11.2011 / Accepted 26.07.2012}

\abstract
{A first characterization of extrasolar planets by the observational determination of the radius has recently been achieved for a large number of  planets. For some planets, a measurement of the luminosity has also been possible, with many more directly imaged  planets expected in the near future. The statistical characterization of  exoplanets through their mass-radius and mass-luminosity diagram is becoming possible. This is for planet formation and evolution theory of similar importance as the mass-distance diagram.}
{Our aim is to extend our planet formation model into a coupled formation and evolution model. We want to calculate from one single model in a self-consistent way all basic quantities describing a planet: its mass, semimajor axis, composition, radius and luminosity. We then want to use this model for population synthesis calculations.}
{In this and a companion paper, we show how we  solve the structure equations describing the gaseous envelope of a protoplanet not only during the early formation phase, but also during the gas runaway accretion phase, and during the evolutionary phase at constant mass on Gyr timescales. We  improve the model further with a new prescription for the disk-limited gas accretion rate, an internal structure model for the planetary core assuming a differentiated interior, and the inclusion of radioactive decay as an additional heat source in the core.}
{We study  the in situ formation and evolution of Jupiter, the mass-radius relationship of giant planets, the influence of the core mass on the radius and the luminosity both in the ``hot start'' and the ``cold start'' scenario. We put special emphasis on the validation of the model by comparison with other models of planet formation and evolution. We find that our results agree very well with those of more complex models, despite a number of simplifications we make in our calculations.}
{The upgraded model yields the most important physical quantities describing a planet from its beginning as a tiny seed embryo to a Gyr old planet. This is the case for all planets in a synthetic planetary population. Therefore, we can now use  self-consistently the observational constraints coming from all major observational techniques. This is important in a time where different techniques yield constraints on very diverse sub-populations of planets, and where its is difficult to put all these constraints together in one coherent picture. Our comprehensive formation and evolution model should be  helpful in this situation for the understanding of exoplanets. }

\titlerunning{Characterization of exoplanets from their formation I}
\authorrunning{C. Mordasini et al.}

\maketitle

\section{Introduction}
The number of known transiting extrasolar planets or planet candidates has recently increased exponentially, thanks both to ground-based observations (e.g. Gillon et al. \cite{gillondoyle2011}), and to space missions like \textit{CoRoT} (e.g. L\'eger et al. \cite{legerrouan2009}) and \textit{Kepler} (Borucki et al. \cite{boruckikoch2011}).  Combined with radial velocity measurements which yield the mass of the planet, one gets the planetary mass-radius  ($M$-$R$) diagram, which is an observational result of similar importance as the semimajor axis-mass  ($a$-$M$) diagram. 

The latter relation is available through the success of ongoing radial velocity surveys (e.g. Mayor et al. \cite{mayormarmier2011}). It is a  goal of population synthesis models to understand the structure of the $a$-$M$ distribution, due to the multitude of clues it contains for planet formation theory (e.g. Ida \& Lin \cite{idalin2004}; Mordasini et al. \cite{mordasinialibert2009a}). A recent comparison of numerous theoretical and observational results mostly obtained by the radial velocity technique can be found in Alibert et al. (\cite{alibertmordasini2011}) and Mordasini et al. (\cite{mordasinialibert2011a}).

The reason for the important role of the $M$-$R$ diagram which is now available for a statistically significant number of planets is that one can derive the mean density of the planet. This constrains the internal planetary structure which is of central importance to understand the nature (Leconte et al. \cite{lecontechabrier2011}), but also the formation of the planet.   {The formation and evolution of the planetary mass-radius relationship is studied in the companion paper Mordasini et al. (\cite{mordasinigeorgy2011}), hereafter Paper II} (see also Mordasini et al. \cite{mordasinialibert2011}).

Besides transiting planets, also the number of planets detected by direct imaging has increased significantly in the past few years, even though in absolute numbers, much less such planets have been found to date. But already these discoveries, like the planetary system around HR 8799 (Marois et al. \cite{maroismacinthosh2008}, \cite{maroiszuckerman2010})  have triggered numerous theoretical studies regarding the formation of these objects (e.g. Dodson-Robinson et al. \cite{dodsonrobinsonveras2009}; Kratter et al. \cite{krattermurrayclay2010}). Two points about these planets are particularly interesting: Their large semimajor axis $a$ and the fact that we measure the luminosity $L$ of young giant planets at some time $t$. Both quantities are important to understand the formation mechanism (e.g. Marley et al. \cite{marleyfortney2007}; Janson et al. \cite{jansonbonavita2011}). In the near future, more capable instruments  like \textit{SPHERE} at the \textit{VLT} (Beuzit et al. \cite{beuzitfeldt2007}) or \textit{GPI} at Gemini South (McBride et al. \cite{mcbridegraham2011}) and later  \textit{EPICS} at the \textit{E-ELT} (Kasper et al. \cite{kasperbeuzit2008}) will come online.  We can therefore expect that the number of points we can put in the $t$-$L$, the $a$-$L$ and (for cases with an independent, dynamical mass determination) the $M$-$L$ diagram will increase from a handful at the moment to hundreds in a few years, similar to what has happened for the $M$-$R$ diagram in the  past few years.  

This shift from an era of discovery to one of a first physical characterization of extrasolar planets by their radii and luminosities has profound implications for  planet formation theory. Until recently, planet formation models studied mostly giant planets detected by radial velocity measurement, of which only the minimal mass and some orbital elements were known. Now we are confronted with a multitude of different sets of observational data and constraints, each regarding primarily planets of different types: Transiting, close-in planets, most of them with a small radius (as found by \textit{Kepler}, see also Paper II); directly imagined, self-luminous massive young giant planets at  large distances from their host star; and a wide range in masses going from super-Earth to Jovian mass planets at distances varying from close to the star out to several AU  found by RV measurements (e.g. Mayor et al. \cite{mayormarmier2011}).   
 
The final goal of planet formation theory to develop one theory to explain the formation (and evolution) of all these very different planets is a challenging one, and will require many efforts in the coming years.  Nevertheless,  the approach to bring together all these different observational data  in a coherent way (which in itself is a non-trivial task, cf. Wolfgang \& Laughlin \cite{wolfganglaughlin2011}) to constrain planet formation and evolution theories seems to be a promising route. In the end, one is not interested in a theory which can explain  certain types of planets, but fails for other classes.  

In this and the companion paper we take a first step in this direction. We present multiple upgrades of our formation model (introduced first in Alibert, Mordasini \& Benz \cite{alibertmordasini2004}). The most important addition is that we now calculate not only the formation of the planets, but couple formation in a self-consistent way with the subsequent evolution at constant mass once the protoplanetary disk is gone. With this approach, we can now calculate directly from one single model not only the planet's mass and semimajor axis, but also the planet's main physical characteristics like the radius, luminosity, surface gravity, effective temperature as well as the composition in terms of iron and silicates, ices and H$_{2}$/He. 

These basic characteristics are available for a planet at any time during its ``life'' starting as a tiny sub-Earth mass seed in the protoplanetary disk to a mature, billion of years old planet. A direct coupling of the planet's formation and its evolution is necessary, as it is well known that the formation has  important, direct consequences for the evolution, in particular for the luminosity of giant planets at young ages (``cold'' vs ``hot'' start models, cf. Marley et al. \cite{marleyfortney2007}; Spiegel \& Burrows \cite{spiegelburrows2011}). With this model development, we can now compare our simulations directly with observational constraints coming from radial velocity (and microlensing), transits as well as direct imaging.

Other upgrades are the following:  a detailed description for the rate at which gas is accreted by the planet in the disk-limited gas runaway accretion phase, an internal structure model for the solid (iron/silicate and possibly ice) part of the planet, the inclusion of radioactive decay for the luminosity of the core, a new initial profile for the gaseous disk, a new prescription for the photoevaporation of the disk, including external and internal photoevaporation, and finally a realistically low grain opacity for the gaseous envelope (presented in a dedicated work, Mordasini et al. \cite{mordasiniklahr2011}). 

We thus deal in this paper and in Paper II mostly with improvements of the physical description of one planet. In other papers, we addressed upgrades regarding the disk model (Fouchet et al. \cite{fouchetalibert2011}), the migration of low-mass planets (Dittkrist et al. in prep) or the effect of the concurrent formation of several embryos in one disk (Alibert et al. in prep.).

\subsection{Organization of the paper}
The organization of the paper is as follows: In Sect. \ref{sect:combinedmodel} we give a short overview of the model. In Section \ref{sect:modelupgasenve}, we describe the modifications of the computational module that describes the gaseous envelope structure of the planet, extending it to calculate the  structure not only during the pre-runaway formation phase as in our previous models, but also during the gas runaway accretion/collapse phase and the subsequent evolutionary phase after the disk is gone.  In Section \ref{sect:mdotmax} we address a related subject, namely how to calculate the gas accretion rate in the disk-limited regime, i.e. once gas runaway accretion of forming giant planets has started. 

In Paper II, we present further upgrades regarding the planet module, namely a realistic model for the density of the solid core of the planets and the inclusion of radiogenic heating in it. In Paper II we also describe shortly some modifications regarding the protoplanetary disk model.   {Finally, we use  in Paper II the upgraded model in population synthesis calculations to study the formation and evolution of the planetary $M$-$R$ diagram,  the distribution of planetary radii, and the comparison with observational data.}

In the remainder of this first paper, we show specific results obtained with the upgraded model: In Sect. \ref{sect:examplesinsitu} we study the coupled formation and evolution of a Jovian mass planet at 5.2 AU from the Sun. Many of the effects seen during this particular simulation are characteristic for the effects encountered during the formation and evolution of the planets  in a general population synthesis calculation (Paper II). In Sect. \ref{sect:radii} and \ref{sect:luminosities} we discuss our results concerning the radii and luminosities of giant planets, putting special weight on the comparison with other, more complex models. The luminosity of young Jupiters is further addressed in a dedicated  paper (Mordasini et al. in prep.).

\section{Combined model of planet formation and evolution}\label{sect:combinedmodel}
The formation model used here relies on the core accretion paradigm, coupled to  standard models of protoplanetary disk evolution and tidal migration of  protoplanets. 

The basic concept of core accretion is to follow the concurrent growth of an initially small solid core and its surrounding gaseous envelope, embedded in a disk of planetesimals and gas (Perri \& Cameron \cite{perricameron1974}; Mizuno et al. \cite{mizunoetal1978}; Bodenheimer \& Pollack \cite{bodenheimerpollack1986}). Within the core accretion paradigm, giant planet formation happens as a two step process: first a solid core with a critical mass (of order 10 $\mearth$) must form, then the rapid accretion of a massive gaseous envelope sets in. This process is  thought to take typically several million years. Two other processes in the protoplanetary disk happen on a similar timescale: the evolution of the protoplanetary disk itself,  and the orbital migration of the protoplanets due to angular momentum exchange with the gaseous disk. As described in Alibert et al. (\cite{alibertmordasini2005}), we have therefore  coupled a classical core accretion model (very similar to Pollack et al. \cite{pollackhubickyj1996}) to  models describing the latter two processes, using for the disk a standard $\alpha$ model, and prescriptions for the type I and II migration of the protoplanets. While we use  simple, 1D models for the individual processes, we consider the full coupling between them, which alone leads to complex formation scenarios (cf. Mordasini et al. \cite{mordasinialibert2009a}).  

\subsection{Limitations: no accretion after disk dissipation, primordial H$_{2}$/He envelopes only}
From its conception, our model mainly deals with the formation   {and evolution of relatively massive, Neptunian and Jovian planets} even though that the vast majority of planets that actually form in a population synthesis are low-mass, failed core (proto-terrestrial) planets (Mordasini et al. \cite{mordasinialibert2009b}). We currently do not include the giant impact stage (after the gaseous disk has disappeared) during which terrestrial planet acquire their final mass. Therefore, our model is incomplete in the description of the formation of planets less massive than $\sim10 \mearth$ (see Mordasini et al. \cite{mordasinialibert2009a} for a discussion). 

Regarding the long-term evolution, we only consider primordial H$_{2}$/He envelopes for which we assume that the mass remains constant after the protoplanetary disk is gone, neglecting a possible evaporation of the primordial gaseous envelope, as well as the outgassing of a secondary atmosphere. Such atmospheric mass loss for planets close to the host star will be considered in future work. Note that the minimal allowed semimajor axis for model planets is at the moment 0.1 AU, so that no simulated planets are exposed to the very intense radiation field occurring on very tight orbits like for example CoRoT-7b ($a=0.017$ AU), where atmospheric mass loss may be very important (Valencia et al. \cite{valenciaikoma2010}).   

An exact lower boundary in mass down to which the model can be applied is difficult to specify, as we find in agreement with Rogers et al. (\cite{rogersbodenheimer2011}) that already planets of just a few Earth masses can accrete non-negligible amounts of H$_{2}$/He during the presence of the nebula, provided that the grain opacity in the envelope is low, which is probably the case (Movshovitz et al. \cite{mbpl2010}). In Paper II we study the formation and evolution of a close-in super-Earth  ($\sim4\mearth$) planet with a tenuous 1\% primordial H$_{2}$/He envelope.  We find good agreement of the radius with the study of Rogers et al. (\cite{rogersbodenheimer2011}), showing that the evolution also of this kind of low-mass planets can be modeled. 

We describe in this paper only improvements (or the inclusion of new physical effects) relative to the original model described in Alibert et al. (\cite{alibertmordasini2005}), to which the reader is referred to for the remaining description.

\section{Gaseous envelope calculation during the attached, detached and evolutionary phase}\label{sect:modelupgasenve}
We now turn to the most important improvement of the model, which is the ability to calculate planetary envelope structures during the entire formation and subsequent evolution of the planets. In previous models, we calculated the gaseous envelope only during the phase when the planet had a subcritical core mass less than the one needed to trigger gas runaway accretion (about 10 $\mearth$). This is sufficient if one is interested in the mass of the planets only. In order to characterize the planetary structure (and thus to have $R$ and $L$, too), we now include also the gas runaway phase (which is attained by planets becoming supercritical during the lifetime of the nebula) and  the evolution at constant mass after the dissipation of the nebula. This phase is eventually attained by all planets. 

Note that the calculations of the gaseous envelope in all phases allows  to obtain the core mass more accurately: We can now calculate the capture radius for planetesimals $R_{\rm capt}$  in the gas runaway accretion phase. This capture radius is necessary to calculate the core accretion rate $\dot{M}_{Z}$. In previous calculations, we had to extrapolate the capture radius as found in the pre-runaway phase into the runaway phase.  The behavior of $R_{\rm capt}$ is discussed  in Sect. \ref{sect:formphaseradius}. 

\subsection{Structure equations}\label{sect:structureequations}
The gas accretion rate of the planet (in the early phases before runaway, cf. \ref{sect:formationphasemass}) is obtained by solving the one dimensional, hydrostatic planetary structure equations (e.g. Bodenheimer \& Pollack \cite{bodenheimerpollack1986}). These equations are similar to those for stellar interiors, except that the energy release by nuclear fusion is replaced by the heating by impacting planetesimals, which is the dominant energy source during the early formation stage. The other equations are the standard equations (except for the assumption of a constant luminosity within the envelope, as discussed further down) of mass conservation, hydrostatic equilibrium and  energy transfer (e.g. Broeg \cite{broeg2009}): 

\begin{alignat}{2}
\frac{dm}{dr}&=4 \pi r^{2} \rho    &\quad  \quad \frac{dP}{dr}&=-\frac{G m}{r^{2}}\rho   \label{eq:structureeqs1} \\
\frac{dl}{dr}&=0             & \frac{dT}{dr}&=\frac{T}{P}\frac{dP}{dr}\nabla(T,P)         \label{eq:structureeqs2}  
\end{alignat}

In these equations, $r$ is  the radius as measured from the planetary center, $m$ the mass inside $r$ (including the core mass $\mz$), $l$ the luminosity at $r$, $\rho, P, T$ the gas density, pressure and temperature. The quantity $\nabla(T,P)$ is given by
\beq\label{eq:nabla}
\nabla(T,P)=\frac{d\,{\rm ln}\,T}{d\,{\rm ln}\,P}={\rm min}(\nabla_{\rm ad},\nabla_{\rm rad})
\eeq
where the radiative gradient (in radiative zones) in the diffusion approximation is given as 
\beq\label{eq:lrad}
\nabla_{\rm rad}=\frac{3}{64 \pi \sigma G}\frac{\kappa l P}{T^{4} m},
\eeq
where $\kappa$ denotes the opacity (given by Bell \& Lin \cite{belllin1994} and Freedman et al. \cite{freedmanmarley2008}), while $\sigma$ is the Stefan-Boltzmann constant. The adiabatic gradient (in convective zones) $\nabla_{\rm ad}$ is directly given by the equation of state SCvH (Saumon et al. \cite{saumonchabrier1995}).  We thus assume zero entropy gradient convection and use the Schwarzschild criterion to determine whether a layer is convectively unstable. 

{This means that our model builds on two interlinked traditional assumptions about the interiors of (giant) planets. These assumptions are mainly made for simplicity of the models, but are in fact not necessarily correct (Stevenson \cite{stevenson1985}; Leconte \& Chabrier \cite{lecontechabrier2012}): First, it is assumed, as just stated, that the interiors are adiabatic due to efficient large-scale convection. In this case, for Jupiter and Saturn at present time, the degree of super-adiabaticity (which is the fractional degree by which the actual temperature gradient is larger than the adiabatic gradient) is very small ($10^{-8}-10^{-9}$, Leconte \& Chabrier \cite{lecontechabrier2012}). There are however several situations where the assumption of an adiabatic interior may break down (Saumon \& Guillot \cite{saumonguillot2004}). Additionally, at  early stages of the evolution, the planets could be going through regions in  the surface gravity-effective temperature parameter space where the formation of H$_{2}$ leads to a  non-linear response of the atmospheric structure which influences the degree of super-adiabaticity (Baraffe et al. \cite{baraffechabrier2002}). We are currently working on including the mixing length theory (e.g. Kippenhahn \& Weigert \cite{kippenhahnweigert1990}) in order to better quantify this issue (see also Rafikov \cite{rafikov2007}).}

{The second traditional assumption is that the planet consists of a few well separated regions, which are chemically homogeneous. In our model, we assume that the envelope consists purely of hydrogen and helium. In reality (Stevenson \cite{stevenson1985}; Chabrier \& Baraffe \cite{chabrierbaraffe2007}), it is well possible that compositional gradients exist, originating for example from the dissolution of the core (Wilson \& Militzer \cite{wilsonmilitzer2010}), or the destruction of planetesimals in deeper layers (Mordasini et al. \cite{mordasinialibert2005}).  If a stabilizing compositional gradient is present, it is likely that the large-scale convection breaks up into many thin chemically homogeneous convective layers which are separated by narrow, diffusive interfaces with large compositional gradient (Stevenson \cite{stevenson1979}). Due to the large number of diffusive interfaces, the efficiency of heat transport is strongly reduced, resulting in a super-adiabatic temperature gradient. As recently shown by Leconte \& Chabrier (\cite{lecontechabrier2012}), such a semiconvective model is able to reproduce  the observational constraints coming from the gravitational moments  and the atmospheric composition of Jupiter and Saturn. This shows that an adiabatic interior cannot be taken for granted.}

\subsection{A new, simple method for the calculation of the total luminosity and evolutionary sequences}
In a planetary population synthesis model, the evolution of thousands of different planets is calculated, covering an extremely wide parameter range of planetary core and envelope masses, accretion rates and background nebula conditions. We therefore need a stable and rapid method for the numerical solution of these equations. We have therefore replaced the ordinary equation for $dl/dm=-T \partial S/\partial t$  (where $S$ is the entropy) by the assumption that $l$ is constant within the envelope, and that we can derive the total luminosity $L$ (including solid and gas accretion, contraction and release of internal heat) and its temporal evolution by total energy conservation arguments, an approach somewhat similar to Papaloizou \& Nelson (\cite{papaloizounelson2005}) {and Hartmann et al. (\cite{hartmanncassen1997})}. We first recall that $-dE_{\rm tot}/dt =L$ and that in the hydrostatic case, the total energy ({neglecting rotation}) is given as
\begin{align}
E_{\rm tot}&=E_{\rm grav}+E_{\rm int}=-\int_0^M \frac{G m}{r}\ d m +  \int_{M_{z}}^M u\ d m  \label{eq:etot} \\
&=-\xi \frac{G M^2}{2 R}\label{eq:etotparam}
\end{align}
where $u$ is the specific internal energy, $M$ the total mass, and $R$ the total radius of the planet. The integration of the gravitational energy includes the core, for which we assume for simplicity a density which is constant as as a function of $r$. This is strictly speaking not self-consistent, as we assume that the core is differentiated, see Paper II. Then, the binding energy of the core is given as  $-(3/5) G \mz^{2}/R_{core}$. Note that, by differentiating this energy with respect to time, assuming a constant density, the resulting luminosity from the growth of the core is equal to the accretion luminosity of planetesimals falling onto the core with a velocity equal to 0 at infinity. On the other hand, the core does not contribute to the internal energy as we do not consider the thermal evolution of the core. In Eq. \ref{eq:etotparam}  we have introduced a parameter $\xi$, which represents the distribution of mass within the planet and its internal energy content.  This last formula is indeed nothing else than a definition of $\xi$. {For example, a fully convective, nearly isentropic star that can be approximated by a n=3/2 polytrope would have  $\xi=6/7$ (Hartmann et al. \cite{hartmanncassen1997}).}

The quantity $\xi$ can be found for any given structure at time $t$ with the equations above. Then one can write 
\begin{align}
-\frac{d}{dt}E_{\rm tot}&=L=L_{M}+L_{R}+L_{\xi} \\
&=\frac{\xi G M}{R} \dot{M} \ - \  \frac{\xi G M^2}{2 R^2} \dot{R} \ + \ \frac{G M^2}{2 R}\dot{\xi} 
\end{align}
where $\dot{M}=\mdotz+\mdotxy$ is the total accretion rate of solids and gas, and $\dot{R}$ is the rate of change of the total radius. {This equation corresponds to a generalization of Eq. 6 in Hartmann et al. (\cite{hartmanncassen1997}).} All quantities except $\dot{\xi}$ can readily be calculated at time $t$. We now set 
\beq
L \simeq C\left(L_{M}+L_{R}\right). \label{eq:lumi2}
\eeq
The factor $C$ corrects approximately for neglecting the $L_{\xi}$ term, and is obtained in this way: a posteriori, one can calculate the total energy in the new structure at $t+dt$, which gives the exact luminosity as $L_{\rm ex}=-[E_{\rm tot}(t+dt)-E_{\rm tot}(t)]/dt$.  By setting $C=L_{\rm ex}/L$ one obtains the correction factor so that exact energy conservation would have occurred. As an approximation, we then use this $C$ for the next time step. One finds that with this method, the estimated luminosity $L$ and the actual luminosity $L_{\rm ex}$ agree generally very well, provided that $dt$ is small enough\footnote{The timestep $dt$  for a calculation as in Sect. \ref{sect:examplesinsitu} is set to $\sim10^{4}$ years in phase II. During the collapse  phase, $dt$ must be reduced to $\sim$1 year. Afterwards $dt$ can increase again, reaching $\sim$1 Gyr late during the evolutionary phase. The latter two values correspond to about 0.1-1\% of the Kelvin-Helmholtz timescale of the planet at these moments.}. With this prescription, we can thus always calculate the total luminosity at $t+dt$, which is one of the necessary boundary conditions for the envelope calculations, allowing us in the end to construct evolutionary sequences. We will show below that this method leads to results in terms of luminosity and radius evolution which are in very good agreement with traditional calculations based on the entropy like Burrows et al. (\cite{burrowsmarley1997}) or  Baraffe et al. (\cite{baraffechabrier2003}).   

In this method,  the (core) luminosity due to the accretion of planetesimal during the formation phase, but also the release of energy due to the contraction of the core at constant mass - the dominant contribution from it during the evolution phase for giant planets  (Baraffe et al. \cite{baraffechabrier2008}) - is automatically included. To this luminosity we finally add the radiogenic luminosity $L_{\rm radio}$ (see Paper II) to get the total intrinsic luminosity.  Note that with planetary luminosity we always mean the one emitted from the interior, without the contribution from absorbed stellar radiation which makes up for Jupiter today about 40\% of the total flux (Guillot \& Gautier \cite{guillotgautier2009}).

Note that we plan on further improving our scheme to estimate the luminosity at the next timestep, which is possible by considering the contribution from the core and the envelope separately. 

\subsection{Hydrostatic approximation}\label{sect:hydrostaticapprox}
Equation \ref{eq:etot} assumes that the planet is always in hydrostatic equilibrium, in particular also during the rapid contraction phase at the moment the planet detaches from the nebula {(see the left bottom panel in Fig. \ref{fig:jupi})} which we shall call the collapse phase independent of its true nature. To check if this assumption is  self-consistent, we performed an order of magnitude estimate: Using the temporal change of the total radius $v(R)=dR/dt$ (which is of order -10 m/s during the fastest contraction), we have assumed that the contraction of the layers in the interior is homologous. This allows us to estimate the total kinetic energy as
\beq
E_{\rm kin}=\int_{M_{Z}}^M \frac{1}{2} \left(v(R)\frac{r}{R}\right)^{2} d m. 
\eeq 
Comparing $E_{\rm kin}$ with the total energy, we find that $E_{\rm kin}$ is always many orders of magnitude smaller than the total energy during the entire formation and evolution of a giant planet. During the collapse, it is smaller by about 6-8 orders of magnitude (and during the subsequent slow long-term contraction even by about 20 to 30 orders). This very large difference makes one think that the hydrostatic approximation is justified. A caveat is  that first, our check is only looking at  global quantities and, therefore, would not catch if in certain layers of the planet hydrodynamic effects become locally important. Second, one cannot exclude a priori that the usage of the hydrostatic approximation itself forces the evolution of the planet to behave hydrostatically. 

The large security margin of 6 to 8 orders of magnitudes, and the fact that Bodenheimer \& Pollack (\cite{bodenheimerpollack1986}) reached the  same conclusion seems however indeed to indicate that  the evolution does not become a dynamical collapse, but remains just a rapid gravitational contraction, validating our approach.  We must however note that opposite views do exist (Wuchterl \cite{wuchterl1991}).

\subsection{Simplification using $dl/dr=0$}
The method introduced above only yields the total luminosity at the surface, and not $l(r)$ which is necessary to solve the structure equations. We use the simplest  setting, namely that $dl/dr=0$. This seems at first sight not to be a reasonable approximation as it (formally) means that the complete luminosity originates in the core, which is, of course, not the case except for the early phases. One however finds  that the solution of the structure equations is very insensitive to the specific form that $l$ takes in the interior. This results can  be understood due to the following physical reasons: $l$ enters into the remaining three structure equations only in radiative zones of the planet (see Eq. \ref{eq:nabla} and \ref{eq:lrad}). In convective zones, $l$ does not enter at all. {Note that this is only true for the  assumption that convection is strictly adiabatic, which might not be the case, as discussed at the end of Sect. \ref{sect:structureequations}.}

However, significant radiative zones (in terms of contained mass) only exist during the very early phases of formation of the planet (see e.g. Bodenheimer \& Pollack \cite{bodenheimerpollack1986}). But then the dominant part of the luminosity is generated by the accretion of planetesimals by the core, so that $L \approx L_{\rm core}$, or in other words $dl/dr\approx0$, so that our approximation is close to the exact solution. The core luminosity due to the accretion of planetesimals is given as 
\beq
L_{\rm core}=\frac{G \mz}{\rcore} \mdotz
\eeq
where $\mz$ is the mass of the core, $\rcore$ its radius, and $\mdotz$ is the accretion rate of planetesimals. This means that we use the sinking approximation (Pollack et al. \cite{pollackhubickyj1996}).

During the collapse phase and during the long-term evolution when no planetesimals are accreted any more,  $l$ varies   strongly inside the envelope, from a very small value ($L_{\rm radio}$) at the core to a typically much larger total luminosity at the surface {for giant planets} due to the cooling and contraction of the envelope.  However, in this phase  the planet is nearly fully convective (e.g. Bodenheimer \& Pollack \cite{bodenheimerpollack1986}; Guillot \cite{guillot2005}), with a radiative gradient dominant over the adiabatic one by orders of magnitude (Alibert et al. \cite{alibertmousis2005}), so that the radial variation of $l$ is irrelevant for the structure. A radiative layer still exists,  but  only near the very surface of the planet (e.g. Burrows et al. \cite{burrowsmarley1997}), containing negligible amounts of matter ( $\lesssim1$\%, Guillot \& Gautier \cite{guillotgautier2009}). Therefore, they do not contribute significantly to the contractional luminosity which comes from the deeper interior (Kippenhahn \& Weigert \cite{kippenhahnweigert1990}), so that $dl/dr=0$ is again a good approximation in those parts where $l$ matters. This also means that the Schwarzschild criterion, which uses $l$ to decide if a layer is convective or not, remains valid.   

\subsubsection{Tests of the simplification}\label{sect:testsimplidldr0}
Regarding the assumption that $dl/dr=0$, we performed a number of tests. We adopted different prescriptions for the form of $l$ as a function of radius (or mass). First, we (arbitrarily) assumed a linear increase with $m$ from $l(R_{\rm core})=L_{\rm core}$ at the core to the total luminosity at the surface. 

For the second set of tests, we proceed as follows. Knowing the planetary internal structure at two timesteps (say $t$ and $t-dt$), we compute the Lagrangian change in the specific entropy $S(m,t)$. From this, we compute $l(m,t) = - T (S(m,t)-S(m,t-dt))/dt$, where $T$ is  the mean temperature for mass shell $m$ between $t-dt$ and $t$. Then, we compute the planetary internal structure at time  $t+dt$ by scaling this $l(m,t)$  onto the new core and total mass, and the new $L_{\rm core}$ and total luminosity. This rescaled, variable $l(m,t+dt)$ is then used to calculated the structure.

In all cases, one finds only very small differences between these calculations and the simulations with a constant $l$. For example, for the time needed to go to runaway gas accretion in a Pollack et al. (\cite{pollackhubickyj1996}) J1 like calculation (see the appendix \ref{comparisonp96}), one finds differences of 2-5\% by which it takes longer in the  $dl/dr=0$ simulation compared to the variable $l$ case.  This is a difference which is much smaller than those introduced by e.g. the uncertainties regarding the grain opacity or the planetesimal size. For the long-term evolution phase, the differences are virtually vanishing, because the planets are almost fully convective. This large independence of the results on the form of $l$ has also been noticed by others (A. Fortier, personal communication, 2011).

Finally, we have tested the impact of the simplified luminosity calculation during the attached and collapse phase by comparing with simulations based on the solution of the full set of structure equations using the traditional entropy method, obtained by a completely independent model (Broeg \& Benz \cite{broegbenz2011}). {As described in details in their paper, these authors solve the structure equations in the quasi-hydrostatic approximation on a self-adaptive 1-dimensional grid using the classical Henyey method (Kippenhahn and Weigert, \cite{kippenhahnweigert1990}). For a simulation similar to the J1 case discussed in the Appendix \ref{comparisonp96}, we compared the accretion rates of gas and solids, the luminosity, and the radius of the forming planet.} Also in this case, we found very good agreement.

The tests presented before may seem complicated, and one may wonder why we do not simply solve the whole set of equations, replacing Eq. \ref{eq:structureeqs2}  by its standard form $dl/dm=-T \partial S/\partial t$ . The reason is the following: population synthesis requires the computation of thousands of formation models, starting from different initial condition. Such a computation is done in an automatic way, using programs generating the initial conditions, then computing the formation model, and finally compiling the important results (e.g. mass, semimajor axis, radius, and luminosity of the formed planets). In this situation, it is very important that only very few simulations do not converge, since non converging simulations could introduce a bias in the results (e.g. if all simulations leading to massive planets were not to converge, one would predict only low-mass planets). We have therefore chosen not to use the standard Newton-Raphson in order to compute the planetary internal structure, but to use a shooting method, which converges extremely well (less than 0.3 \% of our simulation {in the synthesis in Paper II} have problems to converge using this method). However, the shooting method is possible in practical only if one knows the luminosity profile \textit{a priori}, which is the case assuming a constant luminosity or in the two sets of tests presented above.

\subsection{Cold versus hot start conditions}
Finally one must take into account that a part of the released gravitational energy of newly accreted material is not incorporated into the planetary structure but already lost as $L_{\rm acc}$ in the accretion shock on the planet's surface (Bodenheimer, Hubickyj, \& Lissauer \cite{bodenheimerhubickyj2000}) or in the surrounding circumplanetary disk (Papaloizou \& Nelson \cite{papaloizounelson2005}). For matter falling in free fall onto the planet, the released luminosity at the surface of the planet  is approximately  given by
\beq
L_{\rm acc}=\frac{G M}{R}\mdotxy 
\eeq
where $\mdotxy$ is the gas accretion rate. The same amount of energy is also lost in total if the matter spirals onto the planet through a circumplanetary disk, with the difference that some fraction (half of it for a Keplerian accretion disk) is already lost in the disk.  

We can now consider two limiting scenarios {(see Spiegel \& Burrows \cite{spiegelburrows2011} for intermediate ``warm start'' simulations)}. First, as in Papaloizou \& Nelson (\cite{papaloizounelson2005}) we assume that all energy liberated in the shock is radiated away from the planet, and does not contribute to the luminosity inside the planet's structure. This is the physically likely scenario (Stahler et al. \cite{stahlershu1980}; Chabrier, personal communication, 2010). The opposite extreme is to assume that no energy at all is radiated away at the shock. While physically unlikely, it is of interest to  consider this case: In absence of radiative losses at the shock, our simulations become similar to the so called ``hot start'' calculations. This is because also for these ``hot start'' models, there is no mechanism which can act as a sink of entropy, which is the central role of the radiative losses at the accretion shock. As demonstrated  Fortney et al. (\cite{fortneymarley2005}) and Marley et al. (\cite{marleyfortney2007}),  the smaller initial entropy of models which assume radiative losses at the shock  has very important consequences for the luminosity of young Jupiters (``cold start'' models, see Sect. \ref{sect:coldstartcompmarley}). 

The luminosity within the planetary structure  $L_{\rm int}$  for the two cases is given as 
\beq
L_{\rm int}=\left\{ \begin{array}{ll}
 L-L_{\rm acc} & \textrm{\ \ \ \ \ \  ``Cold start'' }\\
 L& \textrm{\ \ \ \ \ \ ``Hot start (accreting)''}
\end{array}\label{eq:hotcoldl} \right.
\eeq
while an observer would see the total luminosity $L$ coming from both the intrinsic and accretional luminosity\footnote{It is interesting to note that we should have in principle two components: A cooler (IR) component from the internal luminosity of the planet, and a harder one from the accretion shock. This is similar as for accreting stars. The fact that gas runaway accretion can only occur when there is still a sufficiently massive disk, means that the radiation from the planet likely gets strongly reprocessed by the surrounding disk material, which should make the observable signature more complex.}.  We have added the suffix ``accreting'' for our ``hot start'' calculations, in order to distinguish them from traditional ``hot start'' calculations (Burrows et al. \cite{burrowsmarley1997};  Baraffe et al. \cite{baraffechabrier2003}) where one starts with a fully formed planet (i.e. at the final mass) with some high, arbitrary entropy.  It is clear that our treatment is a strong simplification of the exact physics of the accretion shock (e.g. Stahler et al. \cite{stahlershu1980}). As already mentioned by Marley et al. (\cite{marleyfortney2007}), three dimensional, radiation-hydrodynamic calculations resolving the shock physics would be necessary to get more accurate results. We will show in a dedicated paper (Mordasini et al. in prep.) how our results for the luminosities depend on a number of parameters which should be indicative of general trends for the behavior of the luminosity of newly formed giant planets.  

\subsection{{Possible issues for ``hot'' accretion and $dl/dr=0$}}\label{sect:issuehotdldr0}
{The following concern occurs when ``hot" accretion is simulated with the $dl/dr=0$ simplification in the envelope. When the internal energy  of the infalling gas (or a part of it) is incorporated into the growing object rather than radiated away, this can lead to important modifications of the internal structure, as illustrated by several papers studying the effects of accretion onto low-mass stars (Hartmann et al. \cite{hartmanncassen1997}; Mercer-Smith et al. \cite{mercersmith1984}; Prialnik \& Livio \cite{prialniklivio1985}). These studies show that under some circumstances, the heating of the surface layers by the rapid accretion of hot, high-entropy gas can reduce the temperature gradient to a degree that convection stops. This means that a radiative layer develops in the interior, even at considerable depth.}

{Prialnik \& Livio (\cite{prialniklivio1985}) studied in details the reaction of an initially fully convective 0.2 $\msun$ star to the accretion of  ``cold" and ``hot" gas at a wide range of accretion rates. For the case of  ``cold" accretion, they found that independently of the accretion rate, the objects remain fully convective, and contract as they grow, indicating that the $dl/dr=0$ simplification is appropriate in this situation. Such a behavior of the radius is also found in our simulations of giant planet formation in the ``cold" assumption presented in Sect. \ref{sect:coldstartcompmarley}.  If the gas keeps a fraction $\alpha_h$ of $L_{\rm acc}$ (Eq. 12), different scenarios occur, depending on $\alpha_h$, and the gas accretion rate. In this paper, we study the two limiting cases corresponding to $\alpha_h=1$ for completely ``hot" and $\alpha_h=0$ for completely ``cold" accretion (see Eq. 13).}

{Prialnik \& Livio (\cite{prialniklivio1985}) find that if the gas accretion rate is very small ($\mdotxy\lesssim10^{-10} \msun$/yr$\approx3\times10^{-5}\mearth$/yr), the star remains fully convective also for ``hot" accretion. If both the accretion rate and $\alpha_h$ are high, a significant part of the entire star becomes radiative and the entire star expands by a large amount. For intermediate values of $\alpha_h$ and $\mdotxy$ (including planetary accretion rates expected in a protoplanetary disk $\lesssim10^{-2}\mearth$/yr),  $\sim10-30\%$ of the mass of the star becomes radiative, and the outer radius including the newly accreted matter expands, while the matter originally contained in the star contracts. An expansion of the star undergoing ``hot'' accretion is also seen in the simulations of Mercer-Smith et al. (\cite{mercersmith1984}).}

{As discussed by Prialnik \& Livio (\cite{prialniklivio1985}, see also Hartmann et al. \cite{hartmanncassen1997}), these results can be understood by comparing the accretion timescale of  ``hot'' material with the thermal relaxation timescale of the newly accreted material.  If the accretion timescale is short compared to the relaxation timescale, thermal equilibrium cannot be established, which would be needed for the star to remain convective. In the ``hot start'' simulations (Sect. \ref{sect:hotstartcompbb}) we also see that the radii of the planets reinflate during the gas runaway accretion phase, after the initial collapse. These results will be discussed in details in Mordasini et al. (in prep). It is clear that the presence of a deep radiative zone, which we cannot catch with the $dl/dr=0$ simplification could cause strong departures from the results obtained with this assumption.  This would also affect the luminosity and radius of the planets  after formation.}

{The results of Prialnik \& Livio (\cite{prialniklivio1985}) were obtained for spherically symmetric accretion. In reality, most of the matter would probably be accreted via a circumplanetary disks (e. g. Lubow et al. \cite{lubowseibert1999}), so that only a fraction of the planetary surface would be undergoing direct accretion, while the rest would be free to radiate. As for stars, it seems likely that this could be important for an accurate criterion when convection is inhibited by ``hot'' accretion (Hartmann et al. \cite{hartmanncassen1997}). Three dimensional radiation hydrodynamic simulations seem  necessary to clarify the situation. We will investigate this issue in simulations with a variable luminosity in future work. For the moment, this concern means that our results for the luminosity of young giant planets formed with a ``hot start'' (and especially at high gas accretion rates) must be regarded with due caution. }

\subsection{Inflation mechanism}
At the moment, we do not include  any mechanism which could lead to the so called ``radius anomaly'' (Leconte et al. \cite{lecontechabrier2011}) observed for many transiting Hot Jupiter. These planets have radii clearly larger than expected from standard internal structure modeling as performed here. There are several possible physical mechanisms leading to this bloating like tidal heating (Bodenheimer et al. \cite{bodenheimerlin2001}), dissipation of stellar irradiation deep in the interior (Showman \& Guillot,  \cite{showmanguillot2002}), ohmic dissipation (Batygin \& Stevenson, \cite{batyginstevenson2010}) or double diffusive convection (Chabrier \& Baraffe, \cite{chabrierbaraffe2007}). Not all these mechanisms are yet fully understood, and we leave their study for the moment to dedicated works (e.g. Leconte et al. \cite{lecontechabrier2011}). But we note that population synthesis calculations, including the effects of different such mechanisms, could be a fruitful way to understand which one reproduces best the observed radius distribution {(see Paper II for the predicted synthetic radius distribution for $a\geq0.1$ AU)}.

\subsection{Boundary conditions}\label{subsect:boundary}
In order to solve the structure equations, boundary conditions must be specified, which should also provide a continuous transition between different phases. For the formation and evolution of a giant planet, three different fundamental phases must be distinguished: attached, detached, and evolutionary. Lower mass planets  stay in the first phase until the nebula disappears and then directly pass into the evolutionary phase. Five boundary conditions are given, namely the core radius, the luminosity, the total planetary radius (in the attached phase) or the total planetary mass (in the detached and evolutionary phase), and the surface temperature and pressure. The four differential equations with five boundary condition only have a solution for a given total planetary mass (attached phase) or radius (detached and evolutionary phase).

\subsubsection{Attached (or nebular) phase}
At low masses, the envelope of the protoplanet is attached continuously to the background nebula, and the conditions at the surface of the planet  are the pressure $P_{\rm neb}$ and approximately the temperature $T_{\rm neb}$ in the surrounding disk. The total radius $R$  is given in this regime by about the minimum of the Hill radius $R_{\rm H}$ and the Bondi (or accretion) radius $R_{\rm A}$. The gas accretion rate is found by the solution of the structure equations and is given by the ability of the envelope to radiate away energy so that it can contract (i.e. its Kelvin-Helmholtz timescale), so that new gas can stream in (Pollack et al. \cite{pollackhubickyj1996}). Numerically, for the shooting method, one iterates in this phase on the mass of the envelope. We use 
\begin{align}
R&= \frac{R_{\rm A}}{1 + R_{\rm A}/ (k_{\rm liss} R_{\rm H})} &P&=P_{\rm neb} \label{eq:rout}\\
\tau&={\rm max}\left(\rho_{\rm neb}\kappa_{\rm neb}R, 2/3 \right)  &   T_{\rm int}^{4}&=\frac{3 \tau L_{\rm int}}{8 \pi \sigma R^2}  \label{eq:tauattached}\\[4pt]
 T^{4}&=T_{\rm neb}^{4}+T_{\rm int}^{4} & l(R)&=L_{\rm int}. \label{eq:tattached}
\end{align}
The fitting formula for the radius of the planet in Eq. \ref{eq:rout}  from Bodenheimer et al. (\cite{bodenheimerhubickyj2000}) reduces to $R_{\rm A}$ for $R_{\rm A}\ll  k_{\rm liss} R_{\rm H}$ and to  $k_{\rm liss} R_{\rm H}$ for $k_{\rm liss} R_{\rm H}\ll R_{\rm A}$ where
\begin{align}
R_{\rm A}&= \frac{G M}{c_{\rm s}^{2}} &  R_{\rm H}&=\left(\frac{M}{3 \mstar}\right)^{1/3}a.
\end{align}
The sound speed $c_{\rm s}$  in the equation for the Bondi radius is calculated with the background nebula conditions.  As found by hydrodynamic simulations by Lissauer et al. (\cite{lissauerhubickyj2009}),  only the inner part of the material inside the planet's Roche lobe is permanently bound to the envelope of the planet, while the upper part participates in the general gas flow in the protoplanetary disk.  We therefore follow these authors in setting $k_{\rm liss}=1/3$.

Regarding the temperature, Eq. \ref{eq:tauattached}  reflects the fact that the planet has to be hotter in order to radiate into the nebula (Papaloizou \& Terquem \cite{papaloizouterquem1999}). Equation \ref{eq:tattached} finally means that the planet emits energy from the interior and is concurrently embedded in the protoplanetary nebula with a certain temperature, where $T_{\rm neb}$ is the midplane background nebula temperature as given by our disk model. 

\subsubsection{Detached (or transition) phase}\label{sect:boundariesdetached}
The solution of the structure equations together with the specified boundary conditions leads to the well known result that the gas accretion rate starts to increase exponentially once the core has grown to a mass of the order  of 10 $\mearth$ (see Sect. \ref{sect:examplesinsitu}). This means that at some moment,  the gas accretion rate obtained in this way (i.e, through the planets Kelvin-Helmholtz timescale) must become higher than some external maximal possible gas supply. At this point, the planet enters the second phase and contracts to a radius which is much smaller than the Hill sphere radius.  This is the ``detached'' regime of high-mass, runaway gas accretion planets. The planet adjusts its now free radius to the boundary conditions that are given by an accretion shock  for matter falling at free fall velocity $v_{\rm ff}$ from about the Hill sphere radius to the planet's surface. Probably more realistic boundary conditions would be those appropriate for the interface to a circumplanetary disk (Papaloizou \& Nelson \cite{papaloizounelson2005}).  The maximal possible  gas accretion rate $\mdotxy$ is given by how much gas is supplied by the disk and can flow through the gap onto the planet $\dot{M}_{\rm max}$ (e.g. Lubow, Seibert, \& Artymowicz \cite{lubowseibert1999}). The calculation of this rate is specified in Section \ref{sect:mdotmax}.   The boundary conditions are now (cf. Bodenheimer et al. \cite{bodenheimerhubickyj2000}; Papaloizou \& Nelson \cite{papaloizounelson2005})
\begin{align}
  \mdotxy&=\dot{M}_{\rm XY,max}&   v_{\rm ff}^{2}&=2 G M \left(\frac{1}{R}-\frac{1}{R_{\rm H}}\right)\\
      P&=P_{\rm neb}+\frac{\mdotxy}{4 \pi R^{2}} v_{\rm ff}+\frac{2 g}{3 \kappa} & \tau&={\rm max}\left(\rho_{\rm neb}\kappa_{\rm neb}R, 2/3\right) \label{eq:pdetached} \\
    T_{\rm int}^{4}&=\frac{3 \tau L_{\rm int}}{8 \pi \sigma R^2} & T^{4}&=(1-A) T_{\rm neb}^{4}+T_{\rm int}^{4} \label{eq:tdetached}
\end{align}
The pressure  in Eq. \ref{eq:pdetached} has three components: the ambient pressure (which however soon after the start of the collapse becomes very small compared to the other terms), the ram pressure due to the accretion shock $\rho v^{2}$, and the standard Eddington expression for the photospheric pressure due to the material residing above the $\tau=2/3$ surface. In Eq. \ref{eq:tdetached}, $A$ is the albedo assumed to have the same value of 0.343 as Jupiter (Guillot \cite{guillot2005}), $g$ is the gravitational acceleration $G M/R^{2}$, and for the luminosity we still have $l(R)=L_{\rm int}$.  In future work, we will use an albedo which depends on planet parameters (Cahoy et al. \cite{cahoymarley2010}).  Numerically, one now iterates in this (and the next) phase on the radius $R$.

For the ``hot start (accreting)'' models, which (artificially) assume no radiative losses at all at the shock, it is not completely obvious if we should not include the pressure due to the accretion shock. However, we found that the dominant term for $P$ is the photospheric pressure for all accretion rates we considered, and that the results are similar both with and without the ram pressure. For a gas accretion rate of $10^{-2} \mearth$/yr  the photospheric pressure is always a factor 15 or more larger than the ram pressure, while for $10^{-1} \mearth$/yr, it is still larger by a factor of at least about 2. {This dominance of the photospheric pressure over the ram pressure stems from the low opacities at the typically relevant temperatures ($\sim$1500-3000 K) and densities at the surface of a giant planet undergoing gas runaway accretion. This temperature range approximately corresponds to the interval  where grains have already evaporated, but where the atomic and molecular opacities are still low, resulting in low $\kappa\sim0.01$ cm$^{2}$/g even for full grain opacities. (Bell \& Lin \cite{belllin1994}; Freedman et al \cite{freedmanmarley2008}). The dominance of the photospheric pressure is increased even more by the low grain opacity reduction factor $\fopa=0.003$ which is normally used in the simulations. In simulations which are instead made with  a full grain opacity, and $\mdotxy=10^{-1} \mearth$/yr, the ram pressure becomes dominant by a factor of a few for a short time at the beginning of the detached phase.}

\subsubsection{Evolutionary (or isolated) phase}
The last phase starts when the gaseous disk disappears so that the planet evolves at constant mass. During this phase, we use simple gray stellar photospheric boundary conditions in the Eddington approximation and write (Chandrasekhar \cite{chandrasekhar1939}; Guillot \cite{guillot2005}) 
\begin{align}
    P&=\frac{2 g}{3 \kappa} & T_{\rm int}^{4}&=\frac{ L_{\rm int}}{4 \pi \sigma R^2}\\
  T_{\rm equi}&=280\,{\rm K}  \left(\frac{a}{1 {\rm AU}}\right)^{-\frac{1}{2}}\left(\frac{\mstar}{\msun}\right) &  T^{4}&=(1-A) T_{\rm equi}^{4}+T_{\rm int}^{4}\label{eq:tequievo}
\end{align}
and we have still $l(R)=L_{\rm int}$ which now also equals the total luminosity $L$ as $L_{\rm acc}=0$ (Eq. \ref{eq:hotcoldl}). This procedure means that our outer planetary radius $R$ corresponds to the Rosseland mean $\tau=2/3$ surface. For Eq. \ref{eq:tequievo}, we have assumed that the planet is rotating quickly and redistributing the heat from stellar irradiation over its full surface, and  that the stellar luminosity scales as $\mstar^{4}$ which applies for roughly solar-like stars on the main sequence. In future work, we will include boundary conditions which are also accurate for planets close to the parent star undergoing intense irradiation (Guillot \cite{guillot2010}; Heng et al. \cite{henghayek2011}). For the moment we recall that the minimal allowed distance for planets in the model is 0.1 AU where at least for giant planets, irradiation is not yet a very important factor.
 
It is clear that the boundary conditions for the long-term evolution are described in a much simpler way than in the Burrows et al. (\cite{burrowsmarley1997}) or Baraffe et al. (\cite{baraffechabrier2003}) models which employ proper non-gray atmospheres (see Chabrier \& Baraffe \cite{chabrierbaraffe1997}).  We have however found that the general agreement in terms of total luminosity or radius is very good (cf. Sections \ref{sect:radii} and \ref{sect:luminosities}) and certainly sufficient for our purpose of population synthesis. This is in agreement with Bodenheimer et al. (\cite{bodenheimerhubickyj2000}) who also found that his cooling curves using simple Eddington boundaries and those from Burrows et al. (\cite{burrowsmarley1997}) agree very well.

\section{Disk-limited gas accretion rate of the planet in the runaway regime}\label{sect:mdotmax}
The second improvement of the model presented in this first paper regards the planetary gas accretion rate in the gas runaway accretion phase. It occurs for supercritical cores with a mass larger than about 10 $\mearth$. In this phase, the gas accretion rate is  no longer limited by the planet's envelope structure, but it is limited by the rate at which gas can be provided by the disk. The gravitational interaction of the protoplanet and the surrounding  disk determines the accretion rate across of the growing gap (e.g. D'Angelo et al. \cite{dangelodurisen2010}).

The actual gas accretion rate $\dot{M}_{\rm XY}$ must be given by the smaller of the two rates:
\beq
\dot{M}_{\rm XY}={\rm Min}\left(\dot{M}_{\rm Tkh}, \dot{M}_{\rm XY, max} \right)
\eeq
where $\dot{M}_{\rm Tkh}$ denotes the gas accretion rate as found through the solution of the structure equations in the attached phase, which is determined  by the ability of the envelope to radiate away the potential energy of the accreted matter, while $\dot{M}_{\rm XY, max}$ is the disk-limited rate. 

In our previous models, we used for the disk-limited rate simply
\beq
 \dot{M}_{\rm XY, max}=\dot{M}_{\rm disk,equi} =3 \pi  \tilde{\nu} \Sigma
\eeq
which is the viscous accretion rate in a protoplanetary disk in equilibrium (i.e. where the mass flux is constant as a function of distance from the star). This expression however neglects three points: (i) the presence of the planet acting as a mass sink locally brings the disk  out of equilibrium.  (ii) only a fraction of the gas flux through the disk will be accreted onto the planet, while some part will flow past it (Lubow \& D'Angelo \cite{lubowdangelo2006}). (iii) there can be (initially) a local reservoir of gas around the planet that can be accreted faster than on a viscous timescale.

\subsection{Non-equilibrium flux}\label{sect:nonequiflux}
The inner part of a viscous accretion disk evolves  in absence of  a planet  relatively quickly into equilibrium. However, the sucking action of the planet and  photoevaporation at late stages disturbs the purely viscous evolution and brings the disk out of equilibrium.  Then, the mass flux at a distance $r$ from the star is given as 
\beq
\dot{M}_{\rm disk} =3 \pi  \tilde{\nu} \Sigma + 6\pi r \frac{\partial \tilde{\nu} \Sigma}{\partial r}
\eeq
where the second non-equilibrium term becomes dominant close to the planet when it is in runaway gas accretion. In order to model the mass flux onto the planet, we have to consider different flow geometries of the gas in the gas feeding zone around the planet. We look at the outer flux $\dot{M}_{\rm O}$  at $R_{\rm O}=R_{\rm p}+1/2 R_{\rm out}$ and  the inner flux  $\dot{M}_{\rm I}$ at $R_{\rm I}=R_{\rm p}-1/2 R_{\rm out}$. Here, $R_{\rm p}$ is the orbital distance of the planet, and $R_{\rm out}$ is its gas capture radius which is given in Eq. \ref{eq:rout}.

\subsubsection{Flow geometries}
Four different flow geometries have to be considered, where we count a mass flux towards the star as positive. First we can have a situation where both  
$\dot{M}_{\rm O}$ and $\dot{M}_{\rm I}$ are positive. This is the regular situation. It takes place when the planet is inside the radius of maximum viscous couple $R_{\rm MVC}$ (or radius of velocity reversal) At  $R_{\rm MVC}$,  the flow in the disk  changes from inward accretion to outward spreading. Second, we can have a situation where  both $\dot{M}_{\rm O}$ and $\dot{M}_{\rm I}$ are negative. This is the case if the planet is outside $R_{\rm MVC}$.  The third case occurs when the planet is  exactly at $R_{\rm MVC}$. Then, decretion on both sides happens. This could in principle mean that only a very small amount of gas can be delivered to the planet. But since we take into account the local reservoir of gas (Sect. \ref{sect:localreservoir}), this regime has no practical meaning. The fourth flow geometry occurs when gas flows towards the planet from both sides. This regime occurs due the sucking action of the protoplanet at the beginning of runaway gas accretion, and then again near the end of the disk lifetime, when the global fluxes through the disk are small.

For these four geometries, the flow limited maximal gas accretion rate $\dot{M}_{\rm XY,max,F}$ are calculated as 
\beq 
\dot{M}_{\rm XY,max,F} =\left\{ \begin{array}{ll}
\dot{M}_{\rm O} & \textrm{for}\  R_p < R_{\rm MVC} \\
\dot{M}_{\rm I} & \textrm{for}\  R_p > R_{\rm MVC} \\
{\rm Abs}\left(\dot{M}_{\rm I}-\dot{M}_{\rm O}\right) & \textrm{for}\  R_p = R_{\rm MVC} \\
{\rm Abs}\left(\dot{M}_{\rm I}\right)+\dot{M}_{\rm O} & \textrm{for two sided accretion}.
\end{array} \right.
\eeq
For the first two cases, we have assumed that the planet migrates slower than the gas flows. This is due to the fact that planets which are in gas runaway accretion  typically are also in the planet-dominated type II migration regime, where this condition is fulfilled. For the third case, we consider the net flux, and for the last case, the sum of both fluxes is used.

\subsection{Local reservoir}\label{sect:localreservoir}
A local reservoir of gas which is already in the vicinity of the planet can be accreted at a rate larger than $\dot{M}_{\rm XY,max,F}$, at least for some time. This is relevant at the moment when the planet just passes into runaway gas accretion. Such a higher accretion rate is important for the depth of the so-called ``planetary desert'' (Ida \& Lin \cite{idalin2004}): The higher the possible accretion rates, the lower the number of planets of intermediate masses (ca. 10-100 $\mearth$, see the discussions in Mordasini et al. \cite{mordasinialibert2009a} and Mordasini et al. \cite{mordasinimayor2011}). 

In order to calculate the accretion rate allowed by the local reservoir we use an approach directly based on the local gas surface density, and not on the fluxes. Following D'Angelo \& Lubow (\cite{dangelolubow2008}) and Zhou \& Lin (\cite{zhoulin2007}), we assume a spherical, homogenous, unimpeded accretion flow for a planet with a cross section $\sigma_{\rm cross}$ in a gas of density $\rho \approx \Sigma/H$ (taken as the mean over the gas feeding zone) and a relative velocity $v_{\rm rel}$,
\beq
\dot{M}_{\rm XY,max, R}=\rho \sigma_{\rm cross} v_{\rm rel}.
\eeq
Gas is captured at some radius $R_{\rm gc}$ so that $\sigma_{\rm cross}\approx \pi R^2_{\rm gc}$. For simplicity, we set $R_{\rm gc}$ equal to the radius defined in Eq. \ref{eq:rout}. The relative velocity is approximately given as $v_{\rm rel}\approx R_{\rm gc} \Omega$.  Therefore, the gas accretion rate in runaway, limited by the local reservoir $\dot{M}_{\rm XY,max,R}$ is approximately given as
\beq
\dot{M}_{\rm XY,max,R}\approx\frac{\Sigma}{H}\Omega R^3_{\rm gc}. 
\eeq
As shown by D'Angelo \& Lubow (\cite{dangelolubow2008}), such a simple description of the gas accretion rate is in fairly good agreement with results of 3D hydrodynamic simulations. It is, however, strictly speaking only valid in the intermediate phase where the planet's Hill sphere $R_{\rm H}$ is still smaller than the disk scale height $H$. The accretion flow is  no more spherical and  homogenous once the planet has grown so massive that $R_{\rm H}>H$. Instead it has a more complex geometry due to the tidal interaction of the planet and the disk, leading to gap formation and accretion streamers (see Zhou \& Lin \cite{zhoulin2007}).   However, we find that   $\dot{M}_{\rm XY,max,R}$ is larger than $\dot{M}_{\rm XY,max,F}$ and, therefore, the relevant quantity  only during an intermediate phase just after the start of the disk-limited gas accretion phase (as illustrated in Fig. \ref{fig:mdotmax1} below). In this phase, planets have typically still masses so small that $R_{\rm H}<H$, so that the simple prescription is valid.

The final accretion rate in runaway is calculated as 
\beq
\dot{M}_{\rm XY,max}=k_{\rm lub}{\rm Max}\left(\dot{M}_{\rm XY,max,R},\dot{M}_{\rm XY,max,F}\right)
\eeq
where we have introduced a factor $k_{\rm lub}$ which takes into account that some gas flows past the planet. Following  Lubow \& D'Angelo (\cite{lubowdangelo2006}) we use values of $k_{\rm lub}$ of 0.75 to 0.9. Finally, we also make sure that not more gas is accreted than the amount present in the feeding zone $M_{\rm feed}$, i.e. that $\dot{M}_{\rm XY,max}<M_{\rm feed}/{\rm dt}$, which is however found to be automatically fulfilled with the above equations.

\subsection{Example for the disk-limited gas accretion rate}
\begin{figure}
\begin{center}
\includegraphics[width=\columnwidth]{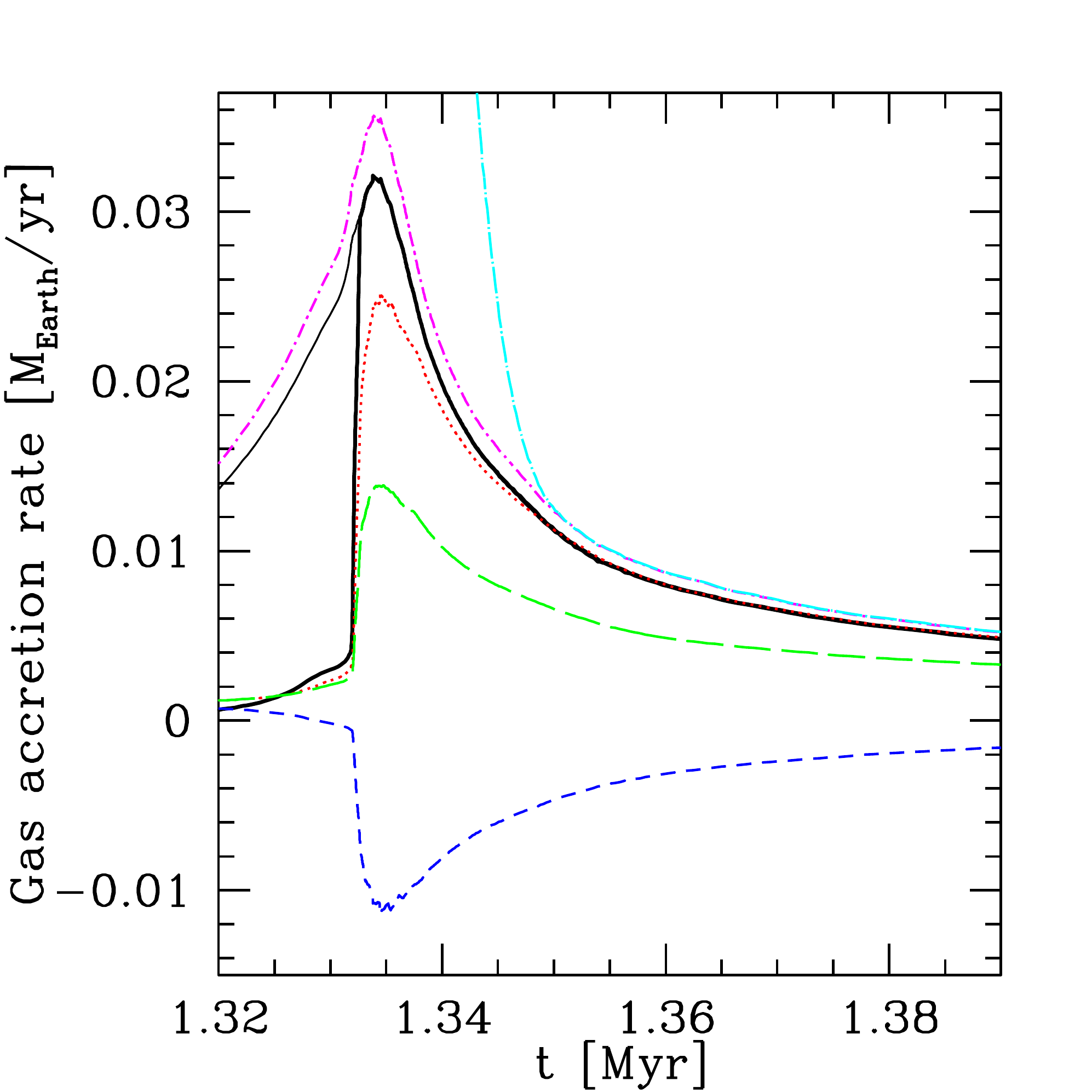}
\caption{Gas accretion rates as a function of time of a forming giant planet, and accretion rate in the surrounding protoplanetary disk at the moment of transition into gas runaway accretion. The  thick black solid line is the accretion rate of the planet $\dot{M}_{XY}$. The thin solid black line is the disk-limited rate,    $\dot{M}_{\rm XY,max}$. The red dotted line is $\dot{M}_{\rm XY,max,F}$, the (non-equilibrium) flow limited maximal accretion rate, while the rate limited by the local reservoir  is shown by the magenta dashed-dotted line ($\dot{M}_{\rm XY,max,R}$). The  cyan long-dashed-dotted line shows $M_{\rm feed}/{\rm dt}$.  The accretion rate in the disk inside the planet's position, $\dot{M}_{\rm I}$, is the blue dashed line while the accretion rate in the disk outside of the planet's position $\dot{M}_{\rm O}$, is the green long-dashed line.}\label{fig:mdotmax1}
\end{center}
\end{figure}

To illustrate the calculation of the gas accretion rate in runaway described in the last section, we show in Fig. \ref{fig:mdotmax1} an example. The figure shows gas accretion rates of a growing giant planet, and accretion rates in the disk during the phase where the planet starts gas runaway accretion. The planet is at about 10 AU in a gas and solid rich disk. The parameter  $k_{\rm lub}$ is set to 0.9, and the viscosity parameter $\alpha$ in the disk is 0.007.

The thick solid black line shows the accretion rate of the planet $\dot{M}_{XY}$. The thin solid black line shows the disk-limited accretion rate,  $\dot{M}_{\rm XY,max}$. Shortly after 1.33 Myr, the accretion rate found by the solution of the structure equations $\dot{M}_{\rm Tkh}$ increases very rapidly, so that  shortly afterwards, the disk-limited rate is reached. In this phase,  $\dot{M}_{\rm XY,max}$ is given as $k_{\rm lub} \dot{M}_{\rm XY,max,R}$. This is because $\dot{M}_{\rm XY,max,R}$ (shown by the magenta dashed-dotted line) is larger than $\dot{M}_{\rm XY,max,F}$ (shown by the red dotted line). We thus find that the local reservoir allows for some time for an accretion rate higher than given by the viscous rate, namely about 0.03 $\mearth$/yr. A typical peak accretion rate of a few times $10^{-2} \mearth$/yr has also been found in other studies (e.g. Lissauer et al. \cite{lissauerhubickyj2009}). Later on, however, (starting at about 1.35 Myrs), we note that both   $\dot{M}_{\rm XY,max,R}$ and  $\dot{M}_{\rm XY,max,F}$ converge on approximately the same value. Also $M_{\rm feed}/{\rm dt}$ (cyan long-dashed-dotted line) takes then approximately the same value. 

From the curves showing the mass flux inside and outside of the planet, $\dot{M}_{\rm I}$ (blue dashed line) and $\dot{M}_{\rm O}$ (green long-dashed line), we can also deduce in which flow geometry we are. Initially, both fluxes are positive, which means that we are in the standard case where the planet is inside of the radius of maximum viscous couple $R_{\rm MVC}$, and the disk is globally flowing towards the star. At about 1.325 Myrs however, we see that  $\dot{M}_{\rm I}$ becomes negative, meaning that the flow geometry changes into the ``two sided accretion'' case. This is a consequence of the accretion onto the planet at a rate higher than the viscous one. Later on (at about 1.6 Myrs, not shown in the figure) the flow geometry changes back into the standard geometry, and  $\dot{M}_{\rm I}$ is then about ten times smaller than $\dot{M}_{\rm O}$. This corresponds to the 10\% of gas that are allowed to flow past the planet. Now, a new equilibrium has established, with the planet acting as an additional sink. 

\section{Results}
We address three different subjects: First, we present a combined formation and evolution calculation for a Jovian mass planet at 5.2 AU. We show the evolution of the most important observable quantities like the  mass, radius and luminosity, but also the evolution of  the central temperature and pressure. The calculation can be compared with the calculations of, e.g.,  Pollack et al. (\cite{pollackhubickyj1996}), Bodenheimer et al. (\cite{bodenheimerhubickyj2000}) or Lissauer et al. (\cite{lissauerhubickyj2009}). Second, we compare the planetary mass-radius relation for giant planets {(see also Paper II)} and the influence of the core mass on the radius as found in our model with several other studies.  Third, we use our model to study the luminosities of young Jupiters. The results for ``cold starts'' can be compared with Marley et al. (\cite{marleyfortney2007}) or Spiegel \& Burrows (\cite{spiegelburrows2011}).

\section{Example of a combined in situ formation and evolution simulation}\label{sect:examplesinsitu}
In this section, we use the extension of the planet module  to calculate an illustrative example of a combined formation and evolution simulation.

These simulations are strongly simplified in comparison to the calculations performed for the complete populations synthesis models (Paper II): In the full model, the moment of the onset of limited gas accretion (and thus the detachment from the nebula), the disk-limited runaway gas accretion rates (Sect. \ref{sect:mdotmax}), as well as the final mass of the planet result in a self-consistent way from the evolution of the gaseous disk (Alibert et al. \cite{alibertmordasini2005}). Here, the evolution of the disk is switched off. Instead, we fix the maximal allowed $\mdotxy$ to some prespecified value, and artificially switch off accretion on a short timescale when the mass of the planet approaches the desired value. Also migration is completely switched off, which is otherwise a central component of the formation model as it can lead to much more complex accretion histories (Mordasini et al. \cite{mordasinialibert2009a}).  But in this way we allow for direct comparison with numerous previous works which used similar assumptions as, e.g., Pollack et al. (\cite{pollackhubickyj1996}), Lissauer et al. (\cite{lissauerhubickyj2009}) or Movshovitz et al. (\cite{mbpl2010}).

In Table \ref{tab:evocalcs}, we list the most important settings used for the simulations. Other settings, like for example a planetesimal radius of 100 km are, the same as in Alibert et al. (\cite{alibertmordasini2005}),  except when mentioned.

\subsection{Grain opacity reduction factor $\fopa$}\label{sect:grainfopa}
A new parameter in our models is the reduction factor of the grain opacity in the envelope relative to the ISM grain opacity, $\fopa$. We have determined this factor in Mordasini et al. (\cite{mordasiniklahr2011}) in the following way:  Movshovitz et al. (\cite{mbpl2010}) coupled for the first time self-consistently a giant planet formation simulation with the concurrent calculation of the grain growth and settling inside the planet's gaseous envelope. They found very low resulting effective opacities, and thus very short durations of  the (plateau) phase II (Pollack et al. \cite{pollackhubickyj1996}). Such grain evolution calculations are currently beyond the scope of our model, {also because they are computationally extremely expensive}. But to still get an approximative representation of this important effect, we have, for different values of $\fopa$, {and the three planetesimal surface densities considered in  Movshovitz et al. (\cite{mbpl2010}),} compared the duration of the phase II as found in our model, with the duration found by Movshovitz et al. (\cite{mbpl2010}). {There is a a very clear relationship between $\fopa$, the planetesimal surface density, and the duration of phase II.} We have found that on the mean {(for the three planetesimal surface densities)}, the duration of phase II becomes the same in both models if we reduce the interstellar grain opacities by a factor 0.003. 

This is  a strong reduction, which is even clearly smaller than the previously studied ``low opacity case'' (Pollack et al. \cite{pollackhubickyj1996}, Hubickyj et al. \cite{hubickyjbodenheimer2005}) which arbitrarily assumed a $\fopa=0.02$. {Note the following two points: First, it is clear that already the calculations of  Movshovitz et al. (\cite{mbpl2010}) are simplified in several aspects like the usage of spherical grains in the opacity calculations instead of aggregates, or the neglect of icy grains. Second, a uniform grain opacity reduction factor cannot reproduce the detailed structure found by Movshovitz et al. (\cite{mbpl2010}) for the opacity as a function of radius in the envelope. It nevertheless leads to envelope structures which are at least similar, especially in the deeper layers. These two points mean that using a uniform grain opacity reduction factor, calibrated via the simulations of  Movshovitz et al. (\cite{mbpl2010}), is a strong simplifications of the actual processes in a protoplanetary envelope. It nevertheless  represents a better approximation than using  arbitrarily reduced values.}  The reader is referred to Mordasini et al. (\cite{mordasiniklahr2011}) for details.

\begin{table}
\caption{Settings for the in situ formation and evolution calculation of Jupiter.}\label{tab:evocalcs}
\begin{center}
\begin{tabular}{lc}
\hline\hline
Quantity & Value \\ \hline
a [AU] & 5.2 \\
 $\sigmas0$ [g/cm$^{2}$]                         & 10          \\                                              
$\dot{M}_{\rm XY,max}$ [$\mearth$/yr]& 0.01 \\
$T_{\rm neb}$ [K]           &150\\
$P_{\rm neb}$ [dyn/cm$^{2}$]           &   0.275     \\ 
Dust-to-gas ratio  &1/70 \\
Initial embryo mass [$\mearth$] & 0.1\\
Migration & not included\\
Disk evolution & not included\\
Planetesimal ejection & included\\
Core density & variable\\
Simulation duration & $10^{10}$ yrs\\
Grain opacity red. factor $\fopa$ &0.003\\ \hline
\end{tabular}
\end{center}
\end{table}

\subsection{Jovian mass planet at 5.2 AU}
For this simulations, we use the same initial planetesimal surface density, disk pressure and temperature as Pollack et al. (\cite{pollackhubickyj1996}). This case has been studied frequently in the literature, and we consider it as our nominal model. The luminosity is calculated as in the ``cold start'' assumption, i.e. we assume that the shock luminosity does not contribute at all to the luminosity inside the planetary structure.

The formation of Jupiter is of particular importance as a benchmark for any planet formation model. This is because for no other giant planet an equally high number of detailed observational constraints exist. In Table \ref{tab:j1comp} in the appendix we list the most important quantities characterizing the planet like the total mass, core mass or luminosity at particular, important moments in time. The data in these tables can be compared with similar data in previous studies mentioned above.

The simulations presented in this section differ from those in Pollack et al. (\cite{pollackhubickyj1996}) in a number of points, making them more realistic. The differences are first that we include planetesimal ejection, second that the core density is variable (Paper II), and third that $\fopa=0.003$, which leads to strongly reduced formation timescale. In the Appendix \ref{comparisonp96} we also present simulations where we set these parameters to the same value as in Pollack et al. (\cite{pollackhubickyj1996}), in order to compare quantitatively  with their simulations. 

\subsubsection{Formation phase: Mass}\label{sect:formationphasemass}
\begin{figure*}
\begin{minipage}{0.5\textwidth}
	      \centering
       \includegraphics[width=0.95\textwidth]{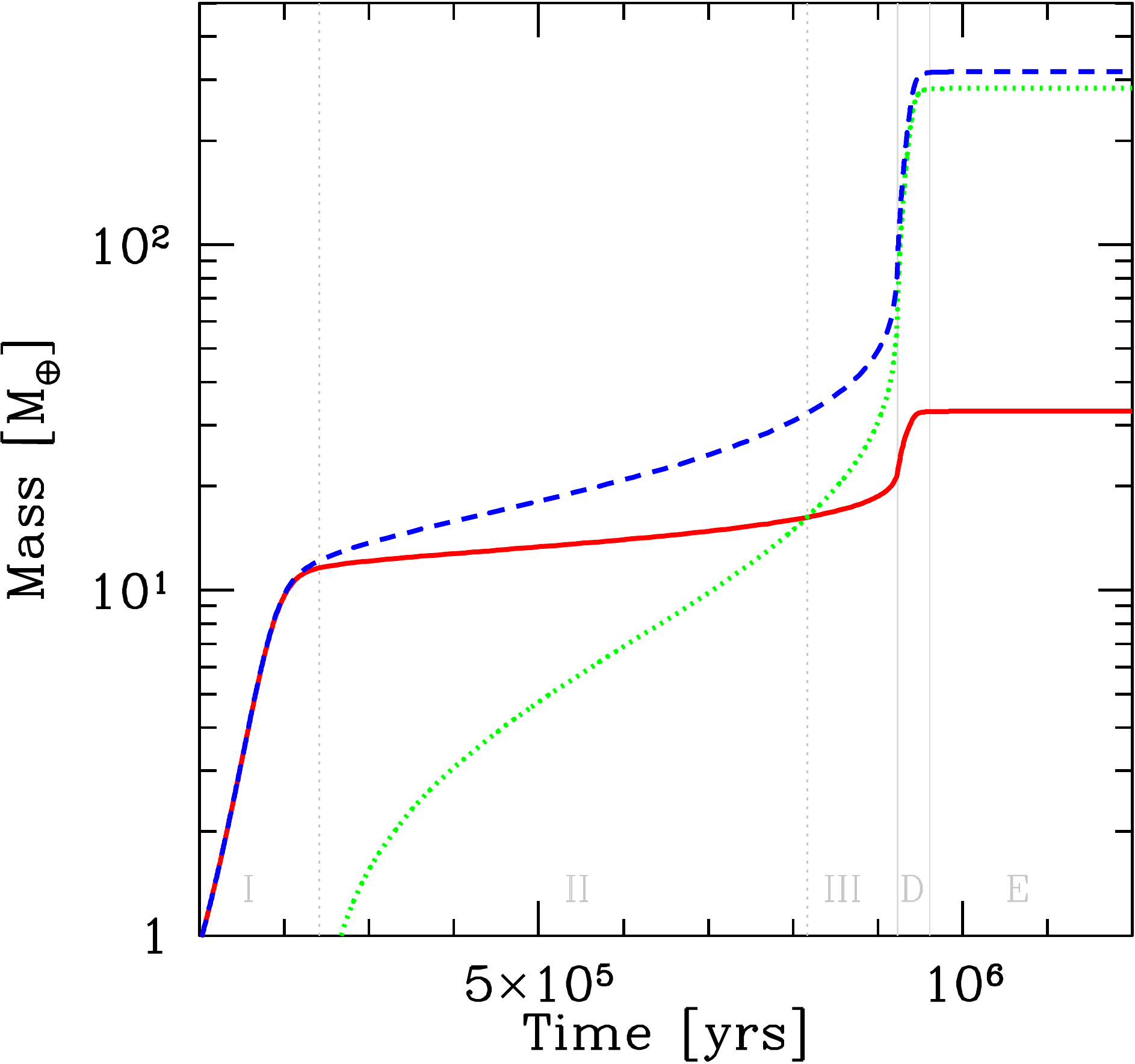}
     \end{minipage}\hfill
     \begin{minipage}{0.5\textwidth}
      \centering
       \includegraphics[width=0.95\textwidth]{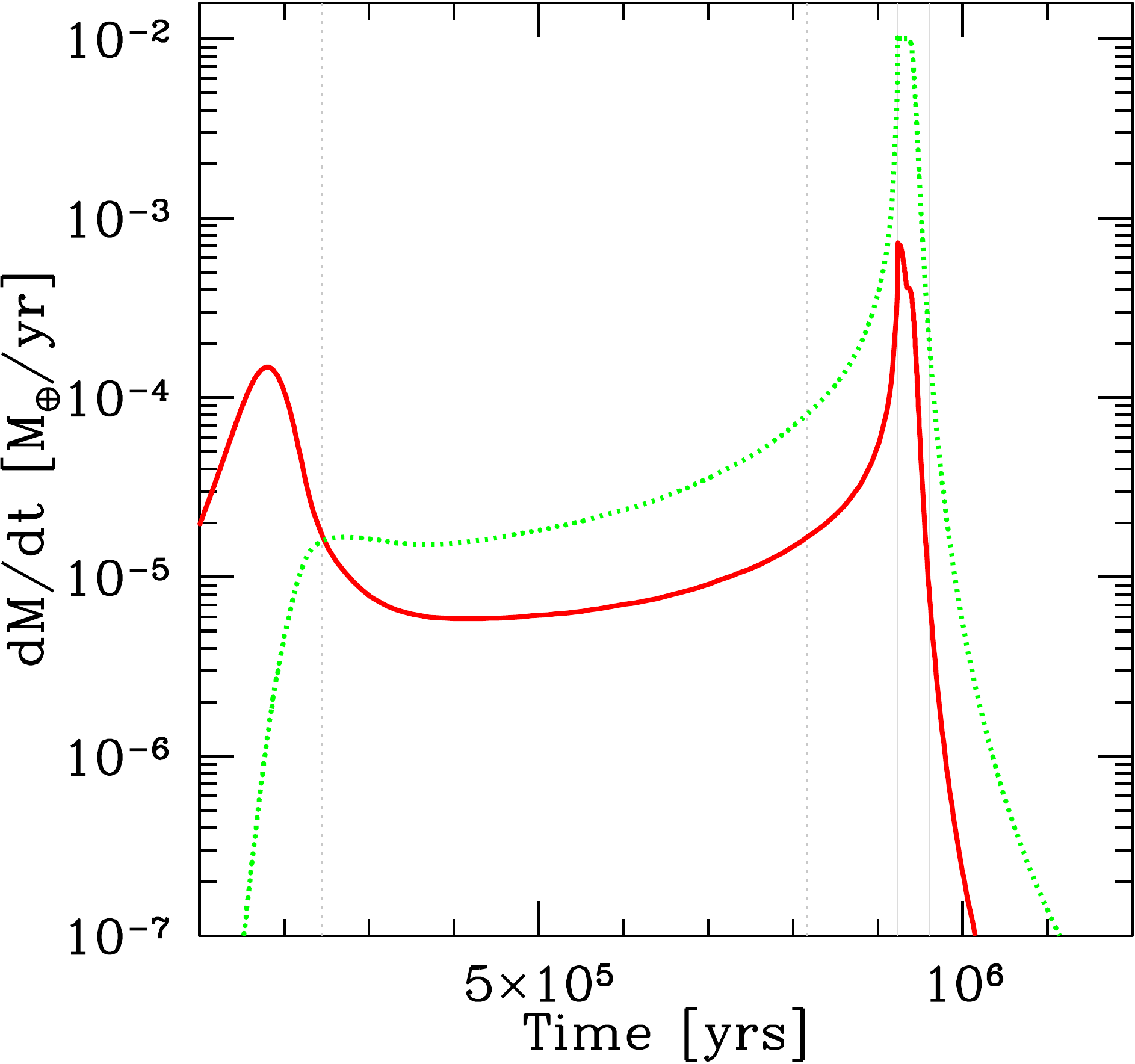}
     \end{minipage}
     \begin{minipage}{0.5\textwidth}
	      \centering
       \includegraphics[width=0.95\textwidth]{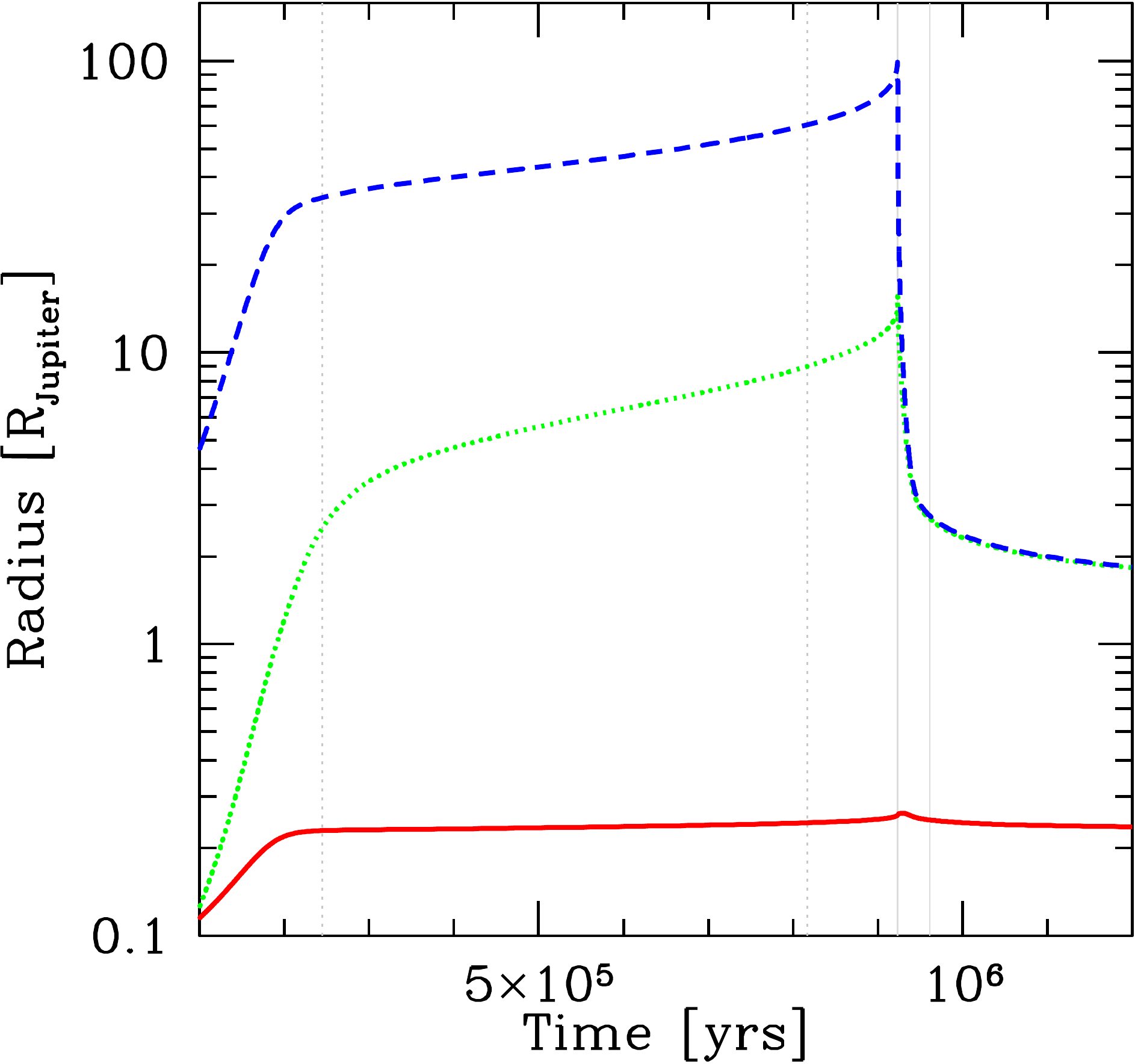}
     \end{minipage}\hfill
     \begin{minipage}{0.5\textwidth}
      \centering
       \includegraphics[width=0.95\textwidth]{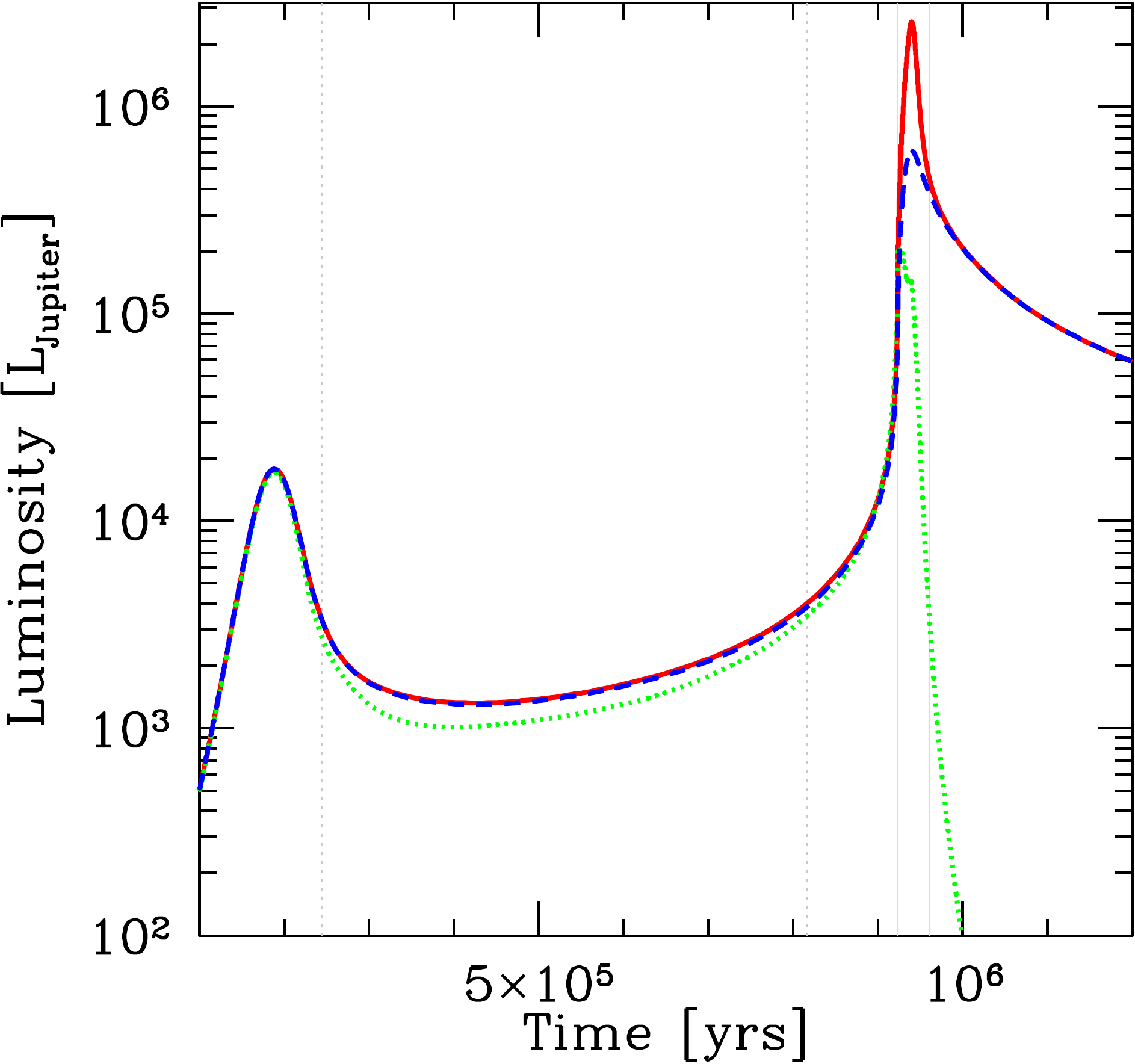}
     \end{minipage}
     \caption{Simulation of the in situ formation of Jupiter at 5.2 AU. Grey, vertical lines show the major phases, labelled  in the top left panel:  I, II, III during the attached phase, D for the detached phase, and E for the evolutionary phase. The top left panel shows the evolution of the core mass (red solid line), the envelope mass (green dotted line) and the total mass (blue dashed line). The top right panel shows the accretion rate of solids $\mdotz$ (red solid line) and of gas $\mdotxy$ (green dotted line). The limiting gas accretion rate is fixed to $10^{-2} \mearth$/yr. The bottom left panel shows the evolution of the core radius $\rcore$ (red solid line), the total radius $R$ (blue dashed line) and the capture radius for planetesimals $\rcapt$ (green dotted line).  The bottom right panel shows the luminosity of the planet in present day intrinsic luminosities of Jupiter ($\lj$=8.7$\times$$10^{-10}\lsun$). The red solid line is the total luminosity $L$, the blue dashed line is the internal luminosity $L_{\rm int}$ and the green dotted line is the core luminosity $L_{\rm core}$.}\label{fig:jupi} 
\end{figure*}

In Fig. \ref{fig:jupi} we show the most important planetary properties as a function of time during the formation and the very beginning of the evolutionary phase once the final mass is reached.

The top left panel shows the characteristic three stages  labelled with I, II and III (Pollack et al. \cite{pollackhubickyj1996}) of such classical in situ calculations (all three occur within the attached phase described in Sect. \ref{subsect:boundary}): First a rapid build-up of the core occurs until the isolation mass is reached, then a plateau phase follows characterized by a slow increase of the envelope mass which allows further core growth, and then the transition to gas runaway accretion is observed. 

The crossover point when the core and  envelope masses are the same is reached relatively  quickly at 0.82 Myrs due to the strongly reduced grain opacity. This duration is in good agreement with Movshovitz et al. (\cite{mbpl2010}). The crossover mass is $M_{\rm cr}=16.3\mearth$.  This is in very good  agreement with Pollack et al. (\cite{pollackhubickyj1996}) with $M_{\rm cr}$=$16.2\mearth$ or Hubickyj et al. (\cite{hubickyjbodenheimer2005})  who obtained $M_{\rm cr}$=$16.1\mearth$ for the same initial planetesimal surface density.

Shortly after the crossover point, the gas accretion rate increases strongly because of beginning runaway accretion, so that it hits the maximal allowed value of  $10^{-2} \mearth$/yr at t=0.923 Myrs. The total mass of the planet is then 86.7 $\mearth$, and $\tau_{\rm KH}$ is as short as $\sim$1640 yrs. At this moment, the second detached phase begins and the collapse of the envelope starts, which is in fact a rapid, but still hydrostatic contraction as discussed in section \ref{sect:hydrostaticapprox} (see also Bodenheimer \& Pollack \cite{bodenheimerpollack1986}). This phase is indicated with the label D in the plot.

At a high gas accretion rate of 0.01 $\mearth$/yr, a mass of 1$\mj$ is approached quickly, so that at 0.935 Myrs we start to artificially ramp down the accretion rates (see top left panel of Fig. \ref{fig:jupi}).  Shortly afterwards, at about 0.961 Myrs, the final mass of the planet is almost reached (99.5\% of the final mass). This marks the beginning of the evolutionary phase, labelled with an E.  
Note that the accretion rate of planetesimals also increases strongly during the runaway phase because of the fast expansions of the planet's feeding zone. It reaches a maximum of about $7\times10^{-4} \mearth$/yr and then starts to decrease already before we ramp it down artificially\footnote{The point where the decreases changes from the ``natural'' one to the artificially forced one is visible in the $\mdotcore$ line in the top right panel as a tiny shoulder. This is due to the  $\tanh$-like function (smoothed step function) we use for the artificial ramp down.}. This is due to the fact that ejection of planetesimals becomes increasingly important as the mass of the planet grows (see Alibert et al. \cite{alibertmordasini2005}), and because the capture radius of the planet shrinks (bottom left panel). The final core mass that is obtained is therefore very similar to the one that is found without stopping the accretion artificially, as becomes clear when looking at  Fig. \ref{fig:tmdiffm}.

The final total mass of the planet is 316.6 $\mearth$, while the final mass of the core is 32.9 $\mearth$. In this paper, we assume that all accreted planetesimals can reach the core. In reality, once the envelope has grown above a certain mass (about 1-2 $\mearth$ for 100 km planetesimals, Mordasini et al. \cite{mordasinialibert2005}), planetesimals cannot penetrate any more to the core of the planet but get destroyed in the envelope. Therefore, the 32.9 $\mearth$ value should rather be associated with the total mass of heavy elements contained in the planet than the core mass. 

{The theoretical estimates for the actual core and total heavy element mass in Jupiter vary. Among other factors, especially uncertainties in the EOS of hydrogen and helium (but also in the efficiency of convection) introduce considerable difficulty in an exact determination of the enrichment. Based on the work of Saumon \& Guillot (\cite{saumonguillot2004}),} the total heavy element mass {was estimated by}  Guillot \& Gautier (\cite{guillotgautier2009}) to lie between 10 to 42 $\mearth$. {Besides classical equations of state like the EOS SCvH which are based on the  free energy minimization method, in the past few years EOS based on ab initio quantum molecular dynamics calculations  have become available. The results for the core and total heavy element mass of Jupiter derived from this method, however, vary: Militzer et al. (\cite{militzerhubbard2008}) find a massive core of $\mcore=16\pm2\mearth$ and $\mz= 20\pm4\mearth$, which is quite different from the result of Nettelmann et al. (\cite{nettelmannbecker2012}) of  $\mcore=0-8\mearth$ and $\mz=28-32\mearth$. Finally, Leconte \& Chabrier (\cite{lecontechabrier2012}) who use again EOS SCvH, but allow for semiconvection due to stabilizing compositional gradients, find high $\mz=41-63\mearth$ as the interiors are hotter. These results demonstrate that the investigations regarding the composition of Jupiter have not yet settled on a definitive conclusion. } 

\begin{figure*}
\begin{minipage}{0.5\textwidth}
	      \centering
       \includegraphics[width=0.95\textwidth]{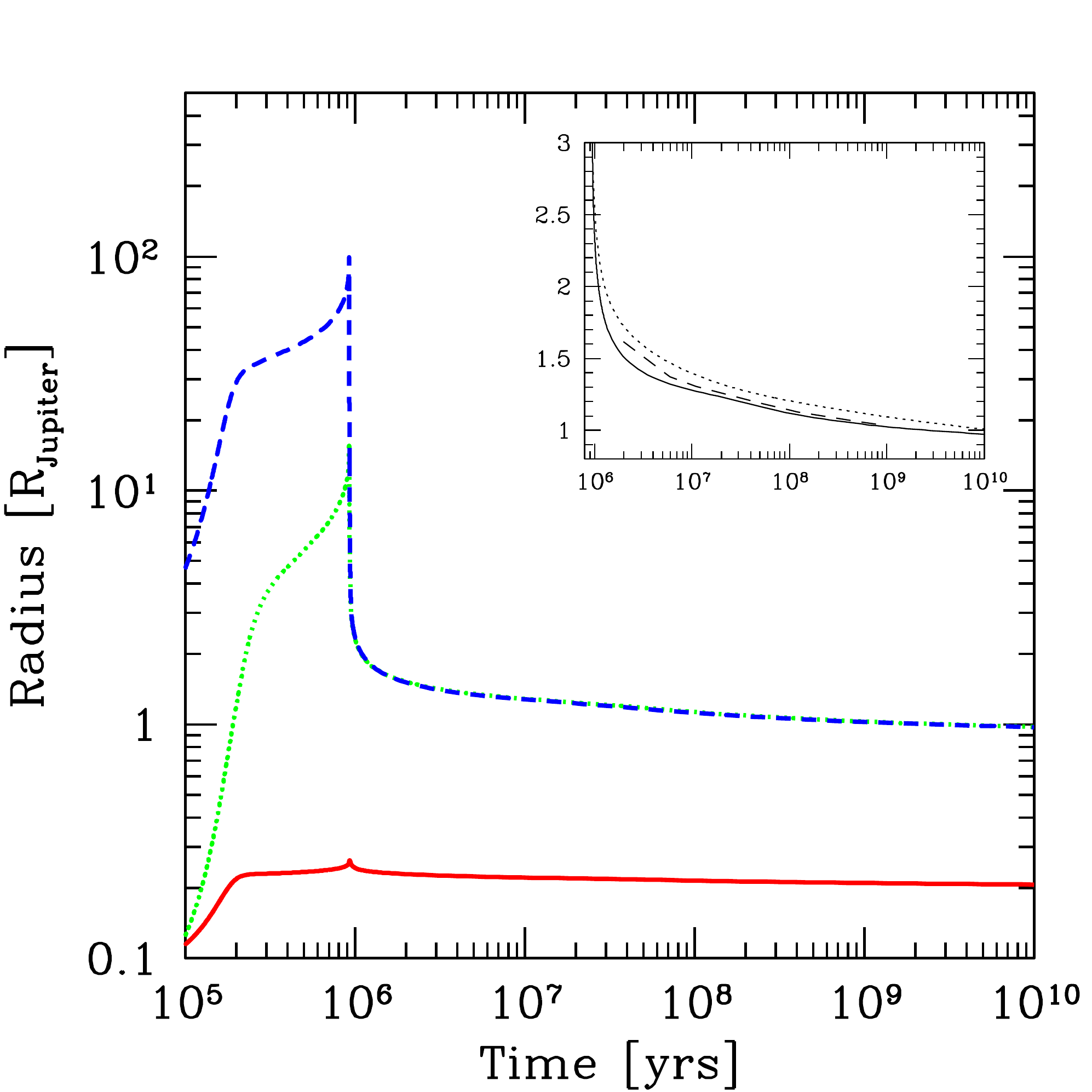}
     \end{minipage}\hfill
     \begin{minipage}{0.5\textwidth}
      \centering
       \includegraphics[width=0.95\textwidth]{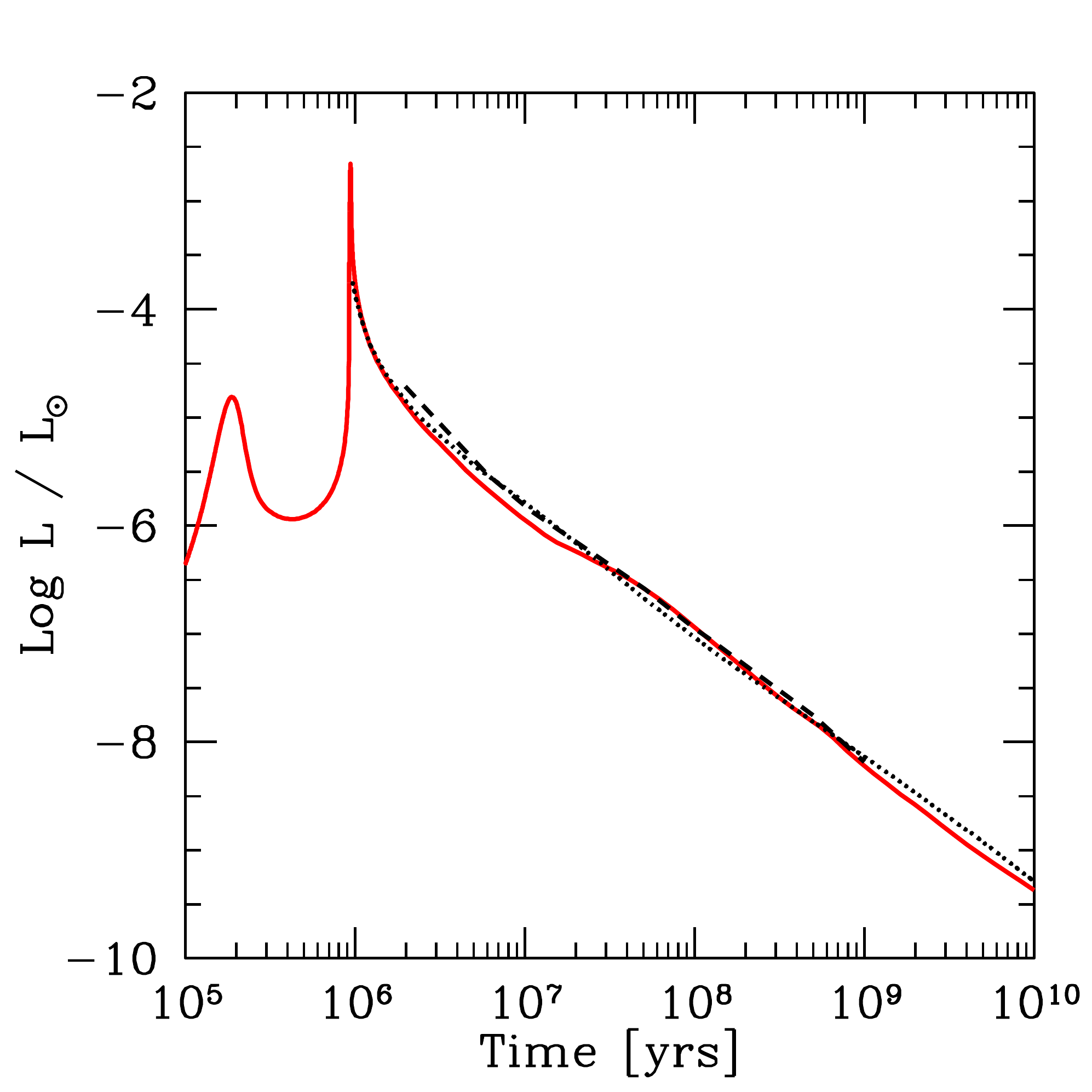}
     \end{minipage}
        \caption{Left panel: Radius $R$ (blue dashed line), core radius $\rcore$ (red solid line) and capture radius $\rcapt$ (green dotted line) as a function of time. The inset figure is a zoom-in onto the late evolution and shows the radius as found in this work (solid line), in Baraffe et al. (\cite{baraffechabrier2003}) (dashed) and Burrows et al. (\cite{burrowsmarley1997}) (dotted). Right panel: Total luminosity $L$ as a function of time (red solid line). The dashed line shows Baraffe et al. (\cite{baraffechabrier2003}) and the dotted one is Burrows et al. (\cite{burrowsmarley1997}).}\label{fig:ltlong} 
\end{figure*}

In our model, the accretion of the core occurs in two steps: about 10-15 $\mearth$ are accreted in phase I and phase II, while the rest is accreted in the gas runaway accretion phase. Then, the mass of the envelope is already high, and planetesimals should be destroyed in the envelope. Therefore, we can estimate that the actual core mass should rather be of order 12 $\mearth$ (this is when the envelope mass reaches 2 $\mearth$), while the remaining 21 $\mearth$ of metals should be dissolved in the envelope, unless they sink to the core-envelope interface (Pollack et al. \cite{pollackhubickyj1996}) .

\subsubsection{Formation phase: Radius}\label{sect:formphaseradius}
The bottom left panel of Fig. \ref{fig:jupi} shows the evolution of the total radius, the core radius and the capture radius. The total radius is initially (during the attached stage) very large, as it is approximately equal to one third of the Hill sphere radius. When the limiting gas accretion rate is hit, the planet detaches from the nebula and collapses to a radius of initially about 2 to 3 Jovian radii. The shape of the curve showing the total radius is very similar to that found by Lissauer et al. (\cite{lissauerhubickyj2009}). 

The capture radius (which is found by integrating the orbits of planetesimals inside the envelope, see Alibert et al. \cite{alibertmordasini2005}) is during phase II much larger than the core radius, namely by a factor between $\sim$10 and 60. This is very important for the core growth, as the capture radius enters quadratically into the growth rate.  After the envelope has started to collapse,  $\rcapt$ becomes equal or slightly less than the total radius. This is due to the fact that in the collapse the envelope structure becomes much more compact. Before the collapse, it is characterized by very large scale heights of thousands to several ten thousands of kilometers, where planetesimals get captured in a soft catch.  Durning the collapse  the planet develops a well defined surface (i.e. a much smaller scale height near the surface) at which planetesimals are captured. 

Looking at the core radius, we note that in the gas runaway accretion phase when the planet grows very quickly in mass, it first increases a bit in size, but then decreases, even though the core grows in this phase by about 20 $\mearth$. The reason is that the external pressure on the  core by the accreted gas increases  so much that the resulting strong increase of the core density over-compensates the mass increase (see Paper II for a dedicated discussion).

The bottom right panel of Fig. \ref{fig:jupi} shows the core luminosity $L_{\rm core}$, the total luminosity $L$ and the luminosity within the planetary structure  $L_{\rm int}$. The difference $L-L_{\rm int}$ is thus the shock/accretion luminosity $L_{\rm acc}$.  The first peak in the curve is due to the rapid accretion of the core, and the second to the combined effects of runaway gas accretion and envelope collapse. During the second peak, for about $2\times10^{4}$ years, the accretional luminosity is larger than $L_{\rm int}$ by up to a factor $\sim$3. For more massive planets discussed below, this ratio can become much larger. The second luminosity maximum occurs at 0.94 Myrs  with $L=2.6$$\times$$10^{6}$ $\lj$ corresponding to $\log(L/\lsun)=-2.65$.  Lissauer et al. (\cite{lissauerhubickyj2009}) find in their simulations with the same limiting gas accretion rate peak luminosities of about $\log(L/\lsun)=-2.35$, which is similar to our value. 

We note that the classical Pollack et al. (\cite{pollackhubickyj1996}) picture with three well defined phases might not be a realistic formation scenario for Jupiter. Several factors including migration (Alibert et al. \cite{alibertmordasini2004}), a slower core growth (Fortier et al. \cite{fortierbenvenuto2007}) or  the opening of a gap in the planetesimal disk (Shiraishi \& Ida \cite{shiraishiida2008}) might suppress in particular phase II. We had two reasons for investigating the classical scenario: First, we want to compare our model with previous studies and second, we are primarily interested in the collapse and evolutionary phase which should be similar  in modified scenarios.

\subsubsection{Evolution phase: Radius  and luminosity}
\begin{figure*}
\begin{minipage}{0.5\textwidth}
	      \centering
       \includegraphics[width=0.94\textwidth]{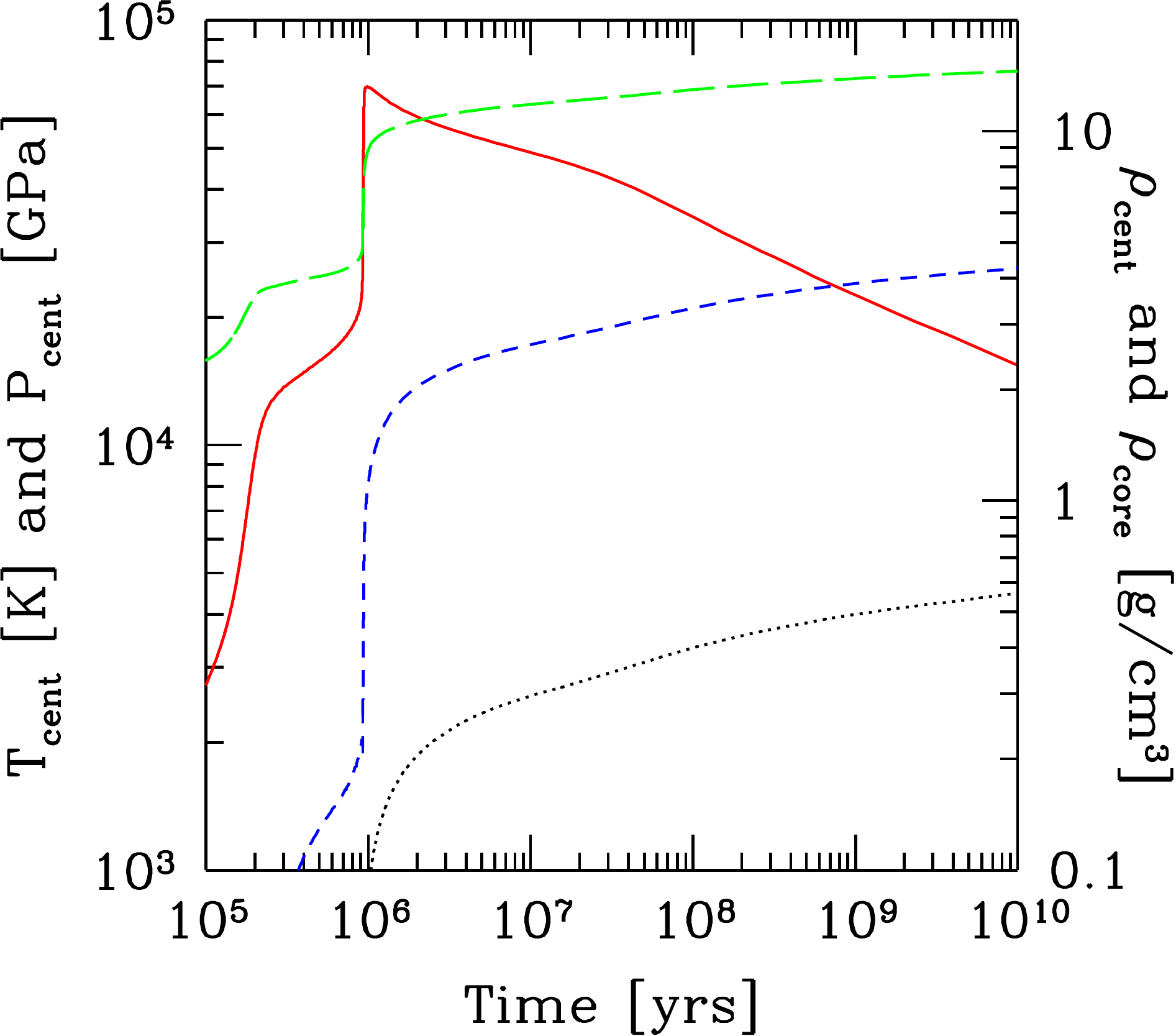}
     \end{minipage}\hfill
     \begin{minipage}{0.5\textwidth}
      \centering
       \includegraphics[width=1.01\textwidth]{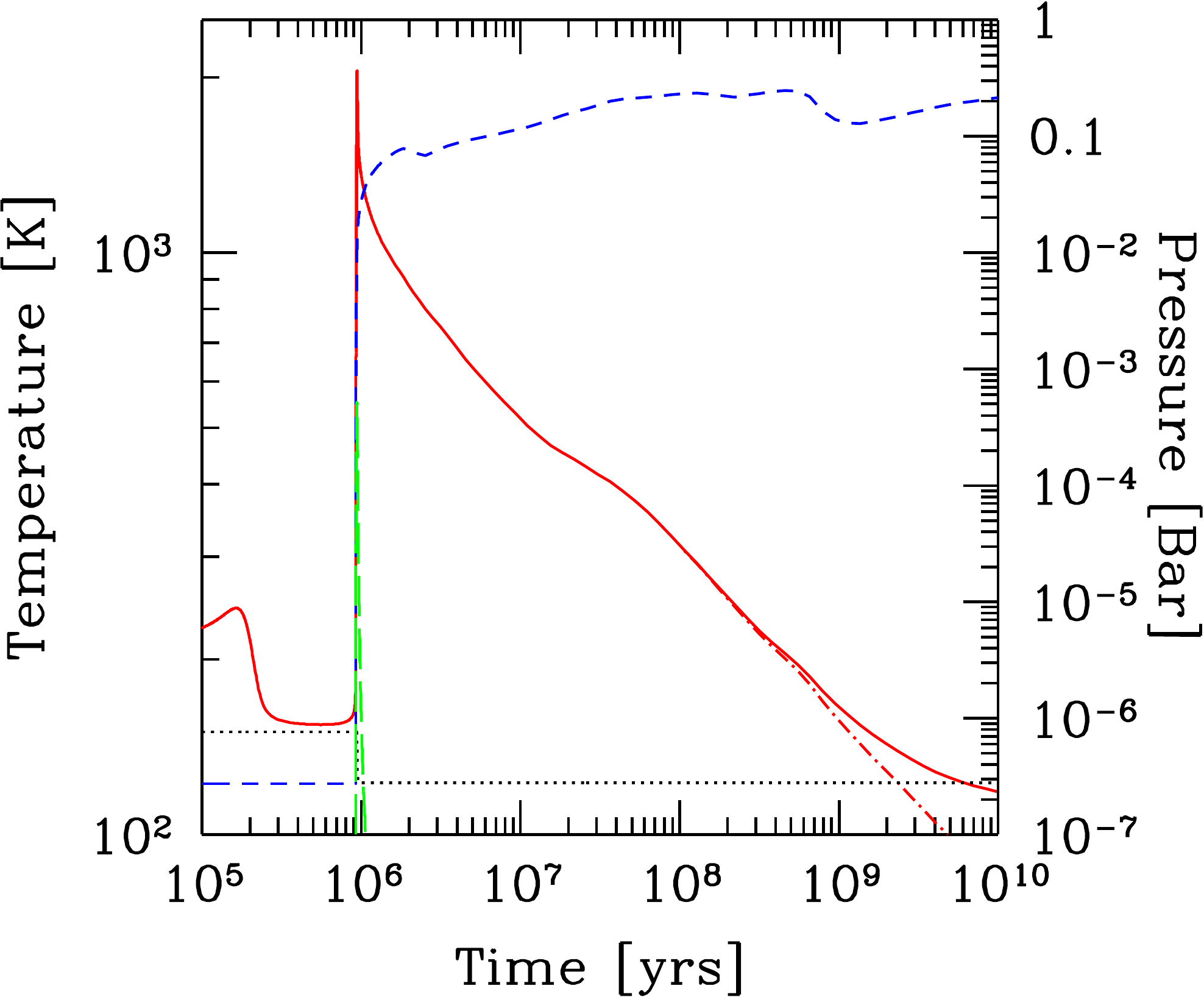}
     \end{minipage}
        \caption{Left panel: Central conditions as a function of time. The red solid line is the temperature at the core-envelope interface, while the dotted black line shows the pressure at this point. These two curves belong to the left axis. The green long-dashed and the blue dashed line are the mean core and the central gas density, respectively. They belong to the right axis. Right panel: Surface conditions as a function of time.  The red solid line is the temperature at the surface, $T_{\rm surf}$, while the dashed-dotted line visible at late times shows the temperature due to the intrinsic luminosity only. The black dotted line is the background temperature. These lines belong to the left axis. The blue dashed line is the (total) pressure $P$. The green long-dashed line shows the contribution from the ram pressure during the gas runaway accretion phase. These lines belong to the right axis.}\label{fig:centsurf} 
\end{figure*}

Figure \ref{fig:ltlong}   shows the radius and luminosity of the planet as a function of time, now including  the long-term evolution over Gigayears when the mass is constant. The evolution now occurs slowly by gradual contraction and cooling. For comparison, the temporal evolution of the radius and the luminosity of a 1 $\mj$ planet as found by Baraffe et al. (\cite{baraffechabrier2003}) and Burrows et al. (\cite{burrowsmarley1997}) is also shown. The temporal zero point in these models has been associated with the time the planet reaches its final mass in our calculations. Baraffe et al. (\cite{baraffechabrier2003}) and Burrows et al. (\cite{burrowsmarley1997}) use non-gray atmospheric boundary conditions, and the standard method based on the $-T \partial S/ \partial t$ term to calculate the evolution. As we can seen, there is good agreement among the models. The differences between our model and the other two ones are of the same order as the mutual differences between Baraffe et al. (\cite{baraffechabrier2003}) and Burrows et al. (\cite{burrowsmarley1997}). Note that the different  internal compositions (mass of heavy elements) have a certain influence on the exact value of  $R$, so that we cannot expect exactly identical results. 

The planet has the following properties after 4.6 Gyrs: A radius of 0.99$\rj$ and a luminosity of 1.13$\lj$ corresponding to $\log(L/\lsun)=-9.01$. At the same age, Burrows et al. (\cite{burrowsmarley1997}) find a radius of 1.11$\rj$ and a luminosity of 1.52$\lj$. Larger values are expected, as the models in Burrows et al. (\cite{burrowsmarley1997}) are core-less models. 

\subsubsection{Central conditions}
In Figure \ref{fig:centsurf}, left panel we show various central conditions during both the formation and evolution phase. With ``central'' conditions we mean here the envelope-core boundary, and not the actual center of the planet. 

The solid red line in the left panel shows the central temperature. It belongs to the scale on the left. Initially, the temperature at the bottom of the envelope is about 3000 K, rising to values between 10\,000 and 20\,000 K during phase II. When gas runaway accretion starts, and the envelope collapses, there is a sharp upturn of $T_{\rm cent}$. The maximal value of 69\,700 K is reached at 0.98 Myrs, afterward it declines again. At an age of 4.6 Gyr, $T_{\rm cent}$ has fallen to about 17\,600 K, which is in good agreement with the estimates of Guillot (\cite{guillot1999}) ranging from 15\,000 to 21\,000 K. 

The second quantity for which the scale is shown on the left is the central pressure, indicated by the black dotted line. It is found to have a value of 4317 GPa at the current age of the Solar System. Guillot (\cite{guillot1999}) gives a value of 4000 GPa.

The other two lines in the figure belong to the y-axis on the right. The green long-dashed line is the mean density of the core, while the blue dashed line is the gas density at the envelope-core boundary. As described in Paper II,  the core  consists of iron, followed by a silicate layer and finally an ice layer, with an ice fraction expected for a body forming outside the iceline. During phase II, $\rho_{\rm core}$ lies between 3 and 4 g/cm$^{3}$. This is close to the typically assumed constant value of 3.2 g/cm$^{3}$ in other formation calculations (e.g. Movshovitz et al. \cite{mbpl2010}).  During the collapse, the external pressure exerted onto the core increases dramatically, so that the the core density increases to about 10 g/cm$^{3}$, leading to the mentioned shrinking of the core size, despite its growth in mass. At 4.6 Gyr, we find $\rho_{\rm core}=14.3$ g/cm$^{3}$.  The central gas density quantitatively follows a similar pattern, starting of course at a smaller value of about 0.1 g/cm$^{3}$ during phase II. At 4.6 Gyr is has reached a fluid like density of  about 4 g/cm$^{3}$.

\subsubsection{Surface conditions}
The right panel of Fig. \ref{fig:centsurf}  shows the conditions at the surface of the planet. The solid red line is the surface temperature $T_{\rm surf}$ calculated as 
\beq
T_{\rm surf}^{4}=T^{4}+ \frac{L_{\rm acc}}{4 \pi \sigma R^{2}},
\eeq 
where $T$ is defined in Eq. \ref{eq:tdetached}. It thus contains the contributions from the intrinsic luminosity, the accretional luminosity which is however important only for a short period (see Fig. \ref{fig:jupi} bottom right panel)  and the absorbed stellar light. When the accretional luminosity is negligible, $T_{\rm surf}=T$, which is the temperature used as the outer boundary condition for the structure. When both $L_{\rm int}$ and $L_{\rm acc}$ are negligible, $T_{\rm surf}=(1-A) T_{\rm equi}$. This happens for a Jovian planet at 5.2 AU after $\gtrsim10$ Gyrs when the intrinsic luminosity of the planet becomes small compared to the incident stellar flux. The background temperature is indicated by the black dotted line.  The background temperature  during the attached and detached phase is equal to the (constant) nebula temperature $T_{\rm neb}=150$ K, while in the evolutionary phase, it is given by the equilibrium temperature with stellar irradiation $T_{\rm equi}$ at 5.2 AU (Eq. \ref{eq:tequievo}). The two temperatures belong to the left scale. One finds that in phase I, $T_{\rm surf}$ can become about a factor 2.5 larger than $T_{\rm neb}$, as $L$ is quite large in this phase. In phase II, $T_{\rm surf}\approx T_{\rm neb}$, in agreement with Papaloizou \& Nelson (\cite{papaloizounelson2005}). During the collapse, there is a strong increase  of $T_{\rm surf}$, which reaches a maximal value of 2050 K.   At the moment  the final mass is reached, $T_{\rm surf}\approx 1350$ K. At the current age of Jupiter, $T_{\rm surf}=126$ K, while the measured value is about 124 K (Guillot \& Gautier \cite{guillotgautier2009}). We see that at late times, the decrease of $T_{\rm surf}$  starts to flatten out as it approaches the constant value of $(1-A) T_{\rm equi}$. In the figure the dashed-dotted red line is $(L_{\rm int}/(4 \pi \sigma R^{2}))^{1/4}$ i.e. the temperature due to the intrinsic flux only. This quantity continues to fall. Note that the nomenclature for the different temperatures is unfortunately not homogenous in the literature: Our quantity $T_{\rm surf}$ corresponds during the evolutionary phase to  $T_{\rm therm}$ according to the definition in Baraffe et al. (\cite{baraffechabrier2003}), but to $T_{\rm eff}$ in Fortney \& Nettelmann (\cite{fortneynettelmann2010}). 

In addition, we show the pressure $P$ at the surface of the planet.  This curve belongs to the right axis. During the attached phase, it is simply equal to the (constant) nebula pressure (Eq. \ref{eq:rout}). Then, at the moment that the planet detaches,  $P$ increases as $P_{\rm edd}=2/3 \ g/\kappa$  becomes large. The dynamic/ram pressure $\rho v^{2}$ also contributes to $P$  in this phase (Eq. \ref{eq:pdetached}).   The ram pressure remains however always 1-2 orders of magnitude smaller than $P_{\rm edd}$ {which is related to the low grain opacity (Sect. \ref{sect:boundariesdetached}).} At late times, after the accretion is over, we see that $P$ has a peculiar shape which is given by opacity transitions.  At 4.6 Gyr, the pressure is about 0.18 bars. 

\subsubsection{Atmospheric pressure-temperature profile}
\begin{figure} 
\begin{center}
\includegraphics[width=\columnwidth]{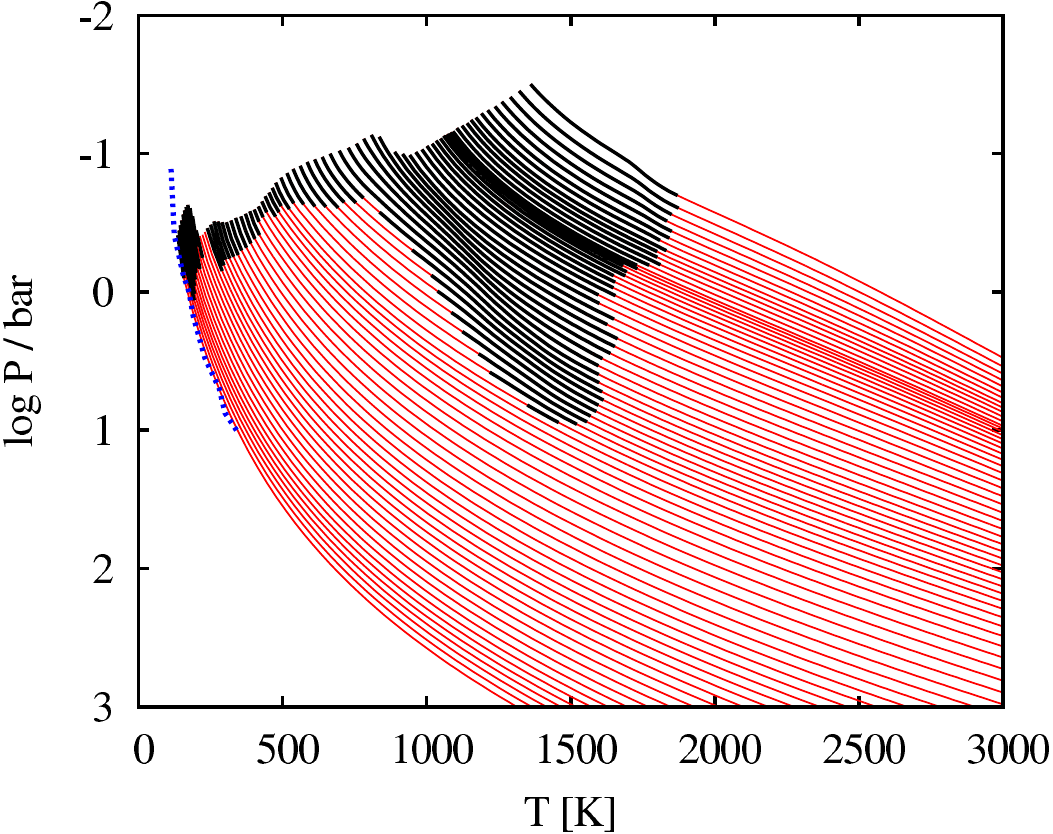}
\caption{Evolutionary sequence of the pressure-temperature profiles near the surface of the planet. The first profile (on the right) corresponds to $t=1$ Myr, while the last one (on the left) is at $t=4.6$ Gyrs. The thick black lines show radiative zones. The blue dotted line is the profile measured by the Galileo space probe (from Lodders \& Fegley \cite{loddersfegley1998}).}\label{fig:ptjupi}
\end{center}
\end{figure}

In Figure \ref{fig:ptjupi} we show the pressure as a function of temperature for the uppermost part of the planet down to a pressure of 1 kbar at different moments in time. The evolutionary sequence starts at the first structure (at the right) at a time of 1 Myr (i.e. shortly after the planet has reached the final mass), while the last structure (at the left) corresponds to  4.6 Gyr. Structures are shown at irregular intervals  between these times.  We recall that the surface of the planet and thus the upper end of the lines corresponds  to the $\tau=2/3$ surface. During the evolutionary sequence, the radius of the planet shrinks from 2.36  to 0.99 $\rj$, which causes an increase of the gravitational acceleration $g$ from 444 at the beginning to 2546 cm/s$^{2}$ at 4.6 Gyrs. In the figure, the thick black lines indicate radiative zones, while the blue dotted line shows the $p-T$ structure measured by the Galileo probe (Lodders \& Fegley \cite{loddersfegley1998}). We find a very good agreement.

The figure can be approximately compared with Fig. 2 in Burrows et al. (\cite{burrowsmarley1997}). These authors use non-gray atmospheres, but indicate the location which corresponds to our outer surface. One must also take into account that they keep the gravity at a constant value of 2200 cm/s$^{2}$ so that we cannot expect to see identical results, especially at early times. But a comparison shows that the two $p-T$ profiles nevertheless have a similar general structure. In both works, there exists for example a detached radiative zone at temperatures around 1500 K.  

In summary we see from the in situ formation and evolution calculation for Jupiter that our combined formation and evolution model, despite the simplifications, leads to a planet with basic properties  in very good agreement with the most important observational constraints coming from Jupiter, and that the new simple method to calculate evolutionary sequences leads to equivalent results as the traditional entropy based method. 

\section{Radii}\label{sect:radii}
We now generalize the results from the last chapter concerning the radius and study the radii of giant planets of different masses and of different compositions. The goal is to validate our model by comparison with existing work (e.g. Fortney et al. \cite{fortneymarley2007} or Baraffe et al. \cite{baraffechabrier2008}). 

We do this by performing the same combined formation and evolution calculations as shown for Jupiter, with the only difference that we vary the initial planetesimal surface density (which will eventually lead to different core masses) and the moment when we terminate the gas accretion, so that we get different total masses. Otherwise, the calculations are identical, which means for example that our calculations apply for planets at a distance of 5.2 AU from a solar like star.

\begin{figure}
\begin{center}
\includegraphics[width=0.95\columnwidth]{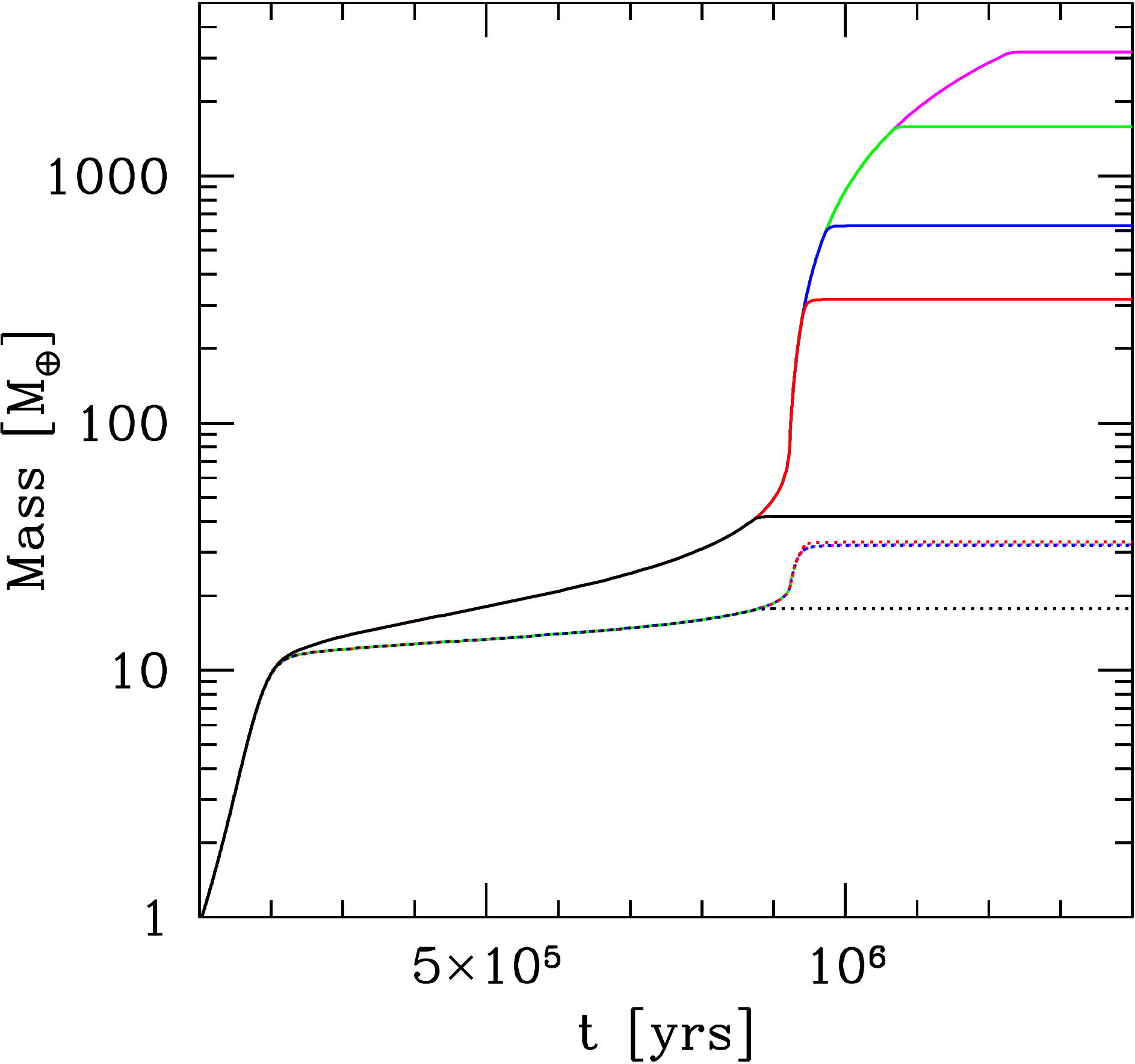}
\caption{Total mass (solid lines) and  core mass (dotted lines) as a function of time for planets with final masses of 0.14, 1, 2, 5 and 10 $\mj$. For the lowest mass planet we shut down accretion before it passes into the runaway gas accretion regime. Note that the four more massive planets have a nearly identical core mass, which is not externally imposed, but a natural consequence of the decrease of the capture radius and the increase of  planetesimal ejection. }
\label{fig:tmdiffm}
\end{center}
\end{figure}

Such calculations are shown in Fig. \ref{fig:tmdiffm}.  The plot shows for planets of final masses of 0.15, 1, 2, 5 and 10 $\mj$ the total and core mass as a function of time. The lines for the 1 Jupiter mass planet are the same as in Sect. \ref{sect:examplesinsitu}. The gas accretion rate in runaway is 0.01 $\mearth$/yr in all cases. Therefore,  more massive planets  reach their final mass later. It is interesting to note that all planets  have a final core mass of about 33 $\mearth$, except for the lowest mass planet (total mass 0.14$\mj=44.5\mearth$) which has a core of about 18 Earth masses. The lower core mass in this case is due to the fact that for this planet, gas and solid accretion are (externally) ramped down before it reaches the gas runaway phase. The nearly identical core mass of all other planets is in contrast not externally imposed. It is rather a natural consequence of the decrease of the capture radius at the moment when the planet collapses (which happens always at the same mass, independent of the final mass) and the ejection of planetesimals which becomes important as the planet grows in mass (Sect. \ref{sect:formationphasemass}). 

\begin{figure}
\begin{center}
\includegraphics[width=\columnwidth]{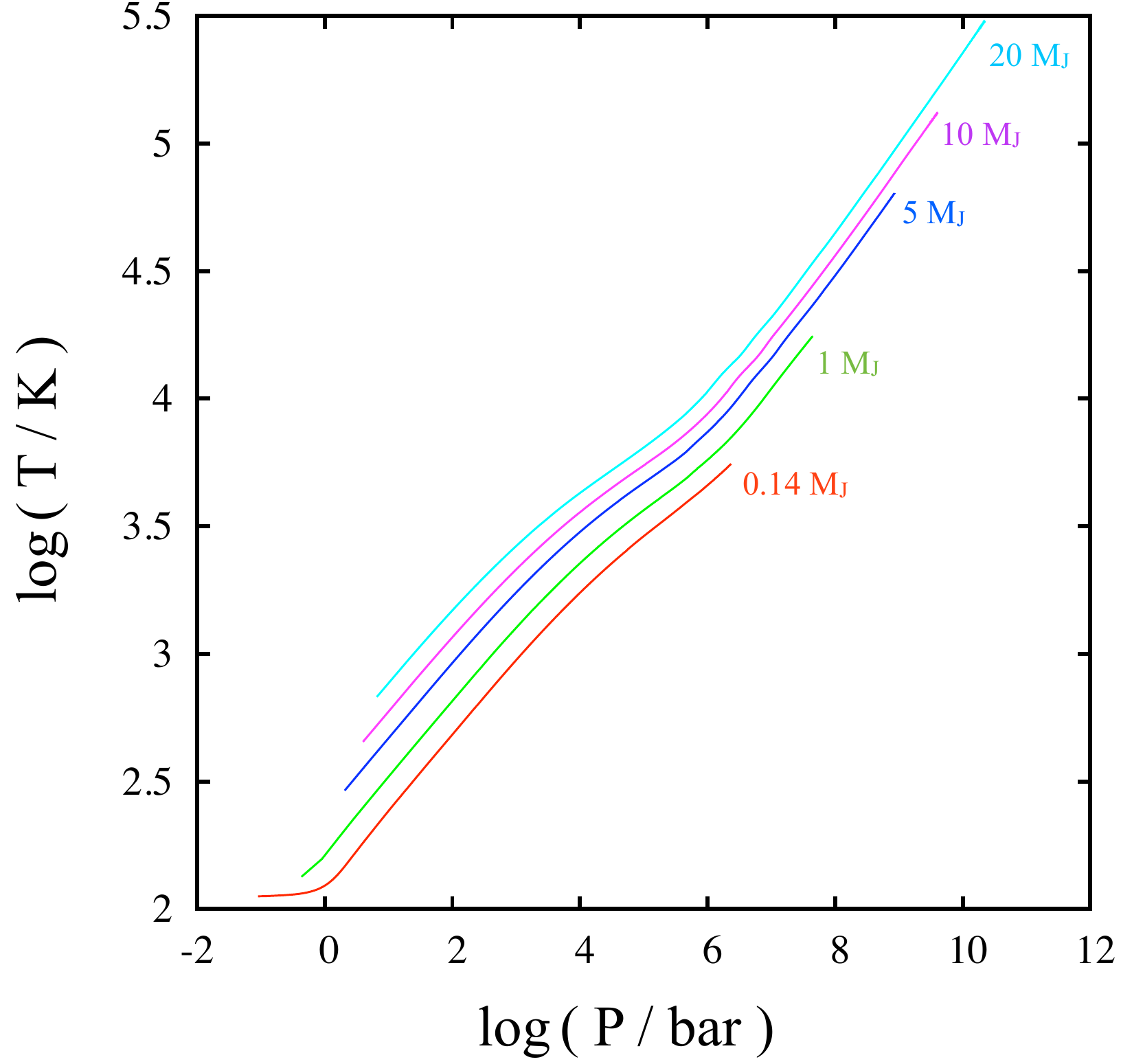}
\caption{Internal temperature-pressure profiles for planets with different masses as labeled in  the figure, at 5.2 AU from the Sun,  and an age between 4 and 5 Gyrs. The core masses of the planets are about 33 Earth masses, except for the 0.14$\mj$ (44.5 $\mearth$) planet which has a core of about 18 $\mearth$. }
\label{fig:pT4p5}
\end{center}
\end{figure}

Figure \ref{fig:pT4p5} shows the internal pressure-temperature structure of these planets (plus also of a 20 $\mj$ planet) at an age of 4 to 5 Gyrs. The left end of the lines corresponds to the surface of the planet, while the right end corresponds to the envelope-core interface. The 0.14 $\mj$ planet has near the surface a significant radiative zone. Otherwise  the planets are nearly fully convective, and characterized by a single adiabat. Near a pressure of about 1 Mbar we see a change in slope which comes from the molecular to metallic transition of hydrogen, also visible in a similar figure in Guillot \& Gautier (\cite{guillotgautier2009}). 

\subsection{Mass-radius relation}
The mass-radius relation of (giant) planets has been studied for a long time (e.g. Zapolsky \& Salpeter \cite{zapolskysalpeter1969}). The interest in the relation lies in its connection to the composition of the planet and the state of  matter in its interior. For recent reviews, see  Chabrier et al. (\cite{chabrierbaraffe2009}) or Fortney et al. (\cite{fortneybaraffe2010}).

The general result qualitatively already found by Zapolsky \& Salpeter (\cite{zapolskysalpeter1969}) is that the $M$-$R$ relationship in the giant planet regime is characterized by a local maximum in $R$. This behavior can be understand with polytropic models with a polytropic index that increases with mass. This change is in turn due to the increasing importance of degeneracy pressure of electrons relative to the classical coulomb contribution of ions with planet mass (e.g. Chabrier et al. \cite{chabrierbaraffe2009}). Another general result is that the radius of giant planets decreases with core mass and increases with  proximity to the star (e.g.  Fortney et al. \cite{fortneymarley2007}; Baraffe et al. \cite{baraffechabrier2008}). 

\begin{figure}
\begin{center}
\includegraphics[width=\columnwidth]{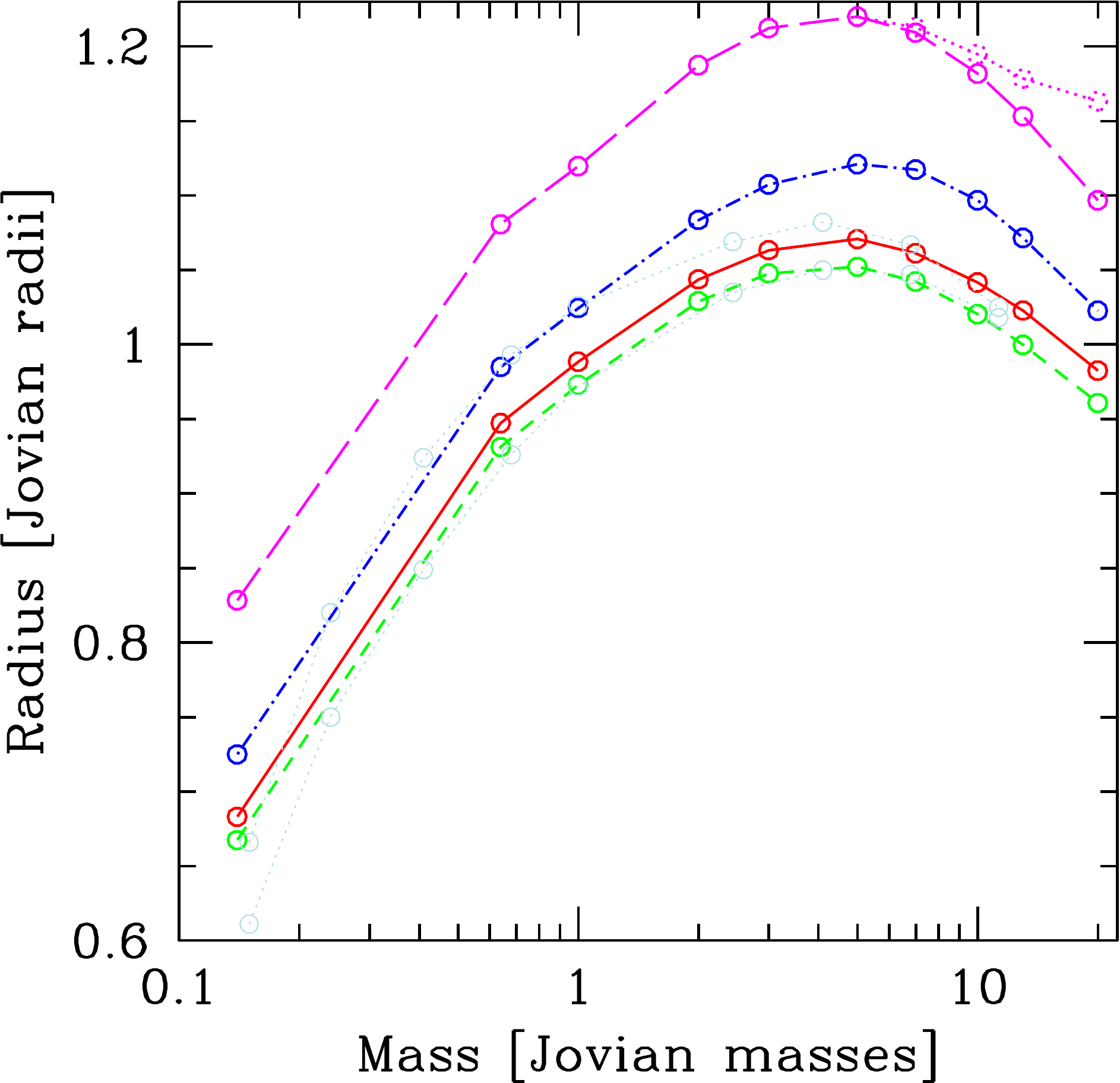}
\caption{Radius of giant planets as a function of mass, for different moments in time. The lines correspond from top to bottom to $t$=0.1 (magenta, long-dashed), 1 (blue, dashed-dotted), 4.6 (red, solid) and 10 Gyrs (green, short-dashed). All planets are at 5.2 AU from the sun, and have a core of $33\mearth$ except for the  0.14 $\mj$ planet which has a core of about 18 $\mearth$. The two dotted, very light blue lines show the result from Fortney et al. (\cite{fortneymarley2007}) for planets with cores equal 25 $\mearth$  at 1 AU (upper line) and at 9.5 AU (lower line), at 4.5 Gyrs. The dotted branch of the magenta line in the top right corner shows the result for ``hot start (accreting)'' simulations. Otherwise  the radii are virtually identical for the ``hot'' and ``cold start'' scenario at these late times. } 
\label{fig:mmr}
\end{center}
\end{figure}

In Figure \ref{fig:mmr} we show the mass-radius relation for planets with masses between 0.14 and 20 $\mj$. This are the same models as in Fig. \ref{fig:tmdiffm} and \ref{fig:pT4p5}. Except for the lowest mass planet with a core of about 18 $\mearth$,  the core mass is approximately 33 $\mearth$. 
The radii are shown at 0.1, 1, 4.6 and 10 Gyrs.  The planets were calculated using the cold start assumption, which however plays only a role at $t=0.1$ Gyrs and for planets more massive than 5 $\mj$. In the top right corner of the figure we show with the magenta dotted line the radius for simulations performed with the ``hot start (accreting)'' scenario. The radii are as expected larger, by about $0.07\rj$ for the most massive planet. For all other cases  the imprint of the formation is lost by $\sim$0.1 Gyrs, and the radii are virtually identical. 

The radii decrease with time, as expected. At all times, there is a maximum of the radius at a mass of about 3-5 $\mj$, in agreement with e.g.  Chabrier et al. (\cite{chabrierbaraffe2009}) or Fortney et al. (\cite{fortneybaraffe2010}). For comparison, we also show  the results from a more detailed model by Fortney et al. (\cite{fortneymarley2007}). The very light blue dotted lines show the radius as found by them at an age of 4.5 Gyrs for planets with a core of 25 $\mearth$ (and thus similar as here) at orbital distances of 1 AU (upper line) and 9.5 AU (lower line). We see that these lines bracket our red line at almost all masses, which is exactly what we expect when the two models yield very similar results. Only for the most massive planets $\gtrsim 10 \mj$ we see that our models yields slightly larger radii, by a few 0.01 $\rj$. Also the lowest mass planet has in our simulations a somewhat larger radius than the counterpart in Fortney et al. (\cite{fortneymarley2007}). This is however not a surprise as it has a smaller core by about 6 $\mearth$, which is relevant due to the small total mass of about 44 $\mearth$ (cf. Fortney et al. \cite{fortneymarley2007}).
 
\subsection{Influence of the core mass}\label{sect:influencecoremass}
The fact that a more massive core  for a given total mass leads to a smaller radius has important implications for the study of  planets. It allows us for example (within certain limits) to infer the bulk composition of transiting extrasolar planets, which in turn constrains formation models (see Paper II). It is for example clear that giant planets formed by the core accretion mechanism must be enriched in heavy elements like it is also the case for Jupiter.  For the competing formation model, the disk instability model, the situation is in contrast less clear (Boley et al. \cite{boleyhelled2011}). 

Another important application is to connect the inferred core masses of transiting  extrasolar planet with their formation environment. In this context, it was found that the core mass of giant planets and the metallicity of their host star is positively correlated (Guillot et al. \cite{guillot2006}; Burrows et al. \cite{burrowshubeny2007}). This is reproduced by planet population synthesis calculations based on the core accretion paradigm (Mordasini et al. \cite{mordasinialibert2009b}). Recent findings (Miller \& Fortney \cite{millerfortney2011}) even indicate that all transiting giant planets contain a minimum amount of 10-15 $\mearth$ of metals, as predicted earlier by the core accretion mechanism (Mordasini et al. \cite{mordasinialibert2011}).

\begin{figure}
\begin{center}
\includegraphics[width=\columnwidth]{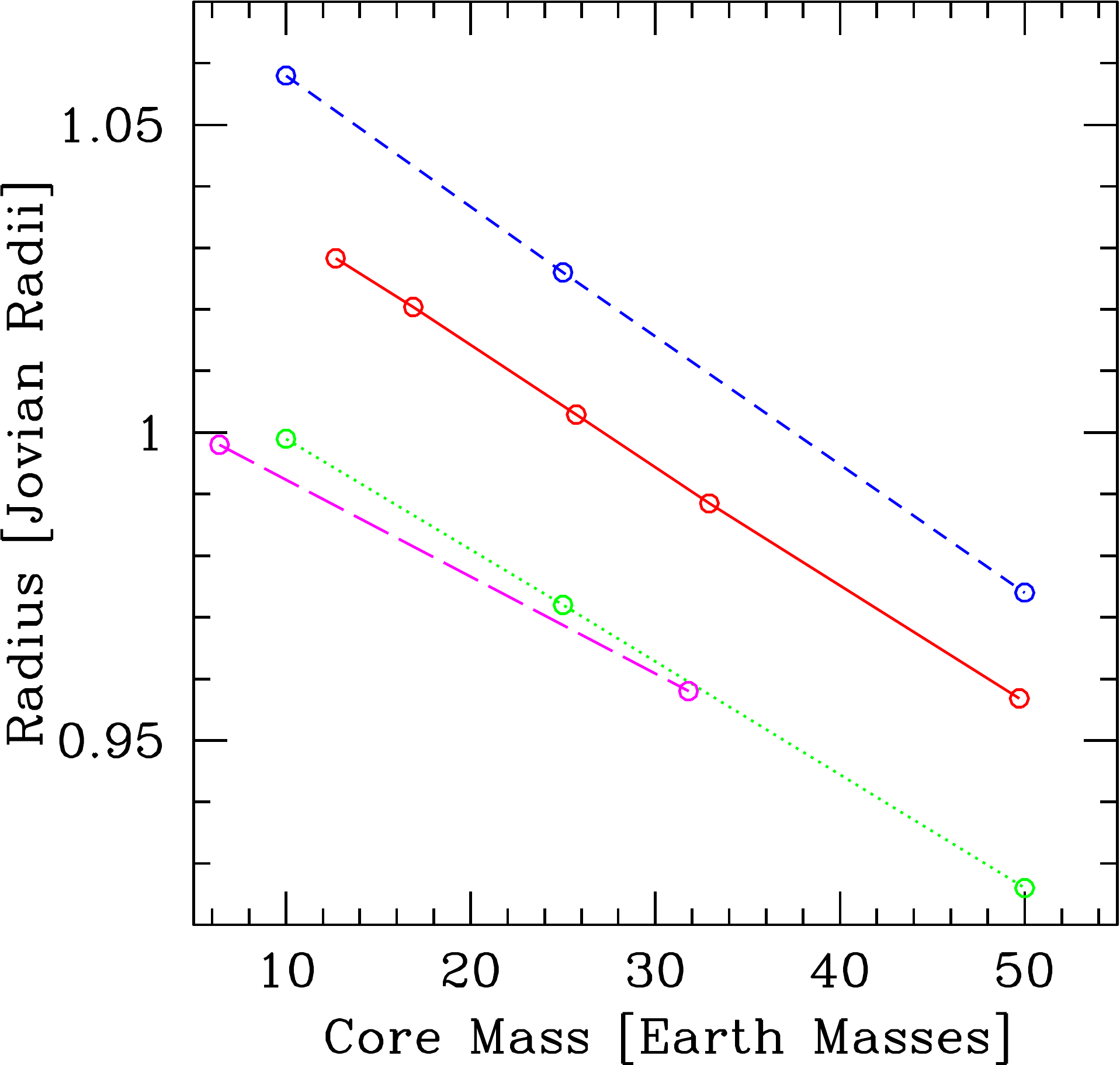}
\caption{Radius of a 1 Jupiter mass planet as a function of the core mass for different models. Red solid line: this work, for $a=5.2$ AU and $t=4.6$ Gyrs.  Blue dashed line: Fortney et al. (\cite{fortneymarley2007}),  $a=1$ AU and $t=4.5$ Gyrs.  Green dotted line: Fortney et al. (\cite{fortneymarley2007}),  $a=9.5$ AU and $t=4.5$ Gyrs. The magenta long-dashed line is from Baraffe et al. (\cite{baraffechabrier2008}), for an isolated object at an age of 5 Gyrs.}
\label{fig:rmcore}
\end{center}
\end{figure}

It is therefore important to study the dependence of the total radius on the core mass  and to compare this with previous studies. Figure \ref{fig:rmcore} shows the radius as a function of core mass for planets with a total mass of 1 $\mj$.  The red solid line shows the result from this work for five planets at an age of 4.6 Gyrs. We see that for an increase of the core mass from $\sim 10$ to $\sim 50$ $\mearth$, there is a reduction of the radius by about 0.07 $\rj$. We also show the results from Fortney et al. (\cite{fortneymarley2007}) and Baraffe et al. (\cite{baraffechabrier2008}). Our result for $a=5.2$ AU are located between the curves for 1 and 9.5 AU from  Fortney et al. (\cite{marleyfortney2007}) at 4.5 Gyrs, with very similar slopes of the curves. From this we conclude that the two models yield very similar results. 

An interesting aspect of the combined formation and evolution calculations is that we can directly relate the radius (respectively the core mass) with the initial surface density of planetesimal during formation, which is not possible for purely evolutionary models. In the current case,    the five different core masses  correspond to planetesimal surface densities of  3.5, 5, 7.5, 10 and 15 g/cm$^{2}$. We must however take into account that there is no unique mapping between planetesimal surface density and core mass, as other factors like migration or the distance where the planet forms also influence this quantity.

\subsection{Comparison and limitations}
We conclude from the different comparisons that our upgraded model allows to get evolutionary sequences  and thus radii with very good accuracy starting with self-consistent initial conditions.

It is however also clear that we have mostly addressed relatively ``simple'' cases of massive, gas-dominated planets at large distances from the star. Already in the solar systems, there are planets which could not be very accurately modeled with a simple model as used here: It is for example well known that for Saturn, He demixing and settling is necessary to explain the measured luminosity (e.g. Hubbard et al. \cite{hubbardguillot1999}). We must also expect that for low-mass, core-dominated planets, possibly with strongly enriched envelopes, we would get less accurate cooling curves and radii, as we do not model the thermodynamics of the core, assume that all solids are in the core (cf. Baraffe et al. \cite{baraffechabrier2008}), and also ignore effects of the composition on the opacity. Other limitations apply for strongly irradiated planets.   These effects will be added in later modifications of our model.

\section{Luminosities}\label{sect:luminosities}
The other fundamental observable quantity apart from the radius which we obtain from our upgraded model is the planetary luminosity. It is obviously central for  direct imaging observations. For transits observations, the discovery of giant planets with radii much larger than expected from standard internal structure modeling was a surprise. Regarding the direct imaging technique, there is on theoretical grounds an open question, too: what is the luminosity of young Jupiters? Currently, two classes of models are discussed in the literature: ``hot start'' models  (Burrows et al. \cite{burrowsmarley1997}; Baraffe et al. \cite{baraffechabrier2003}), and  ``cold start'' models'' (Fortney et al. \cite{fortneymarley2005}, Marley et al. \cite{marleyfortney2007}). In the ``hot start'' model, one assumes as the initial state an already fully formed planet at an (arbitrarily) high entropy state, and thus a large radius and luminosity. In the ``cold start'' scenario one assumes  that the planet is gradually built up by accretion of gas through an accretion shock on the surface of the planet. The radiative losses  of the liberated gravitational energy at the shock lead to a lower entropy, and thus lower luminosity and radius.

\subsection{{Relationship of cold/hot start and core accretion/gravitational instability}}
{The importance of the two different  initial states stems from the fact that the luminosity of young planets is an observable quantity establishing links to the physical mechanisms occurring during the formation. The currently known directly imaged planets seem to be incompatible (Janson et al. \cite{jansonbonavita2011}; Marley et al. \cite{marleysaumon2012}) with the ``cold start'' model of Marley et al. (\cite{marleyfortney2007}), since the observed temperatures and luminosities are higher than predicted by this model. It should be kept in mind that the model of Marley  et al. (\cite{marleyfortney2007}) studies the limiting case of a shock that is 100\% efficient in radiating away the accretional energy (and makes a number of assumptions concerning the formation process, cf. below). This limiting case is studied as detailed, possibly multi-dimensional radiation-hydrodynamic calculations of the shock properties have not yet been conducted. Recently, Spiegel \& Burrows (\cite{spiegelburrows2011}) have therefore studied a broad range of  intermediate ``warm-start'' models showing that this leads to significant differences in the brightness of the planets depending on the mass and spectral band.}

{The related question about  the initial thermodynamic state of  newly formed brown dwarfs and stellar objects has been studied in various works (e.g. Prialnik \& Livio \cite{prialniklivio1985}, Hartmann et al. \cite{hartmanncassen1997}; Baraffe et al. \cite{baraffechabrier2009}; Commer\c{c}on et al. \cite{commerconaudit2011}). These studies indicate that depending on the fraction of accretion luminosity radiated away or absorbed by the protostar, the formation through gravitational collapse can lead to bright,  extended objects ("hot accretion" leading to a "hot start") or faint and compact objects (``cold accretion'' leading to a ``cold start").  The latter is the case if the accretion shock during the formation process is supercritical, i.e. if all the accretion shock energy is radiated away, as found recently by Commer\c{c}on et al. (\cite{commerconaudit2011}) for  accretion on the first Larson core. }

{For giant planets, two formation mechanisms are currently discussed. If  a ``cold'' or ``hot start'' can preferentially be associated with one of the two formation mechanisms,  then the understanding of the luminosity of young Jupiters has important implications for the understanding of giant planet formation as a whole.}  

\subsubsection{Core accretion}
{The first proposed formation mechanism is core accretion, which is the paradigm underlying this work. A number of observational findings indicate that core accretion represents the dominant formation mechanism at least for planets inside a few AU (e.g. Mordasini et al. \cite{mordasinialibert2011a}).  Whether core accretion can also explain the directly imaged, massive planets at orbital distances $\gg10$ AU is the subject of an ongoing debate (e.g. Dodson-Robinson et al. \cite{dodsonrobinsonveras2009}; Kratter et al. \cite{krattermurrayclay2010}). The core accretion, ``cold accretion'' simulations of Marley et al. (\cite{marleyfortney2007}) lead as mentioned to very low initial luminosities, incompatible with the observed values (Janson et al. \cite{jansonbonavita2011}; Marley et al. \cite{marleysaumon2012}).  We will however show in Mordasini et al. (in prep.) that this result (found for a specific set of parameters during the formation process) does not mean that planets formed by core accretion in a ``cold accretion'' scenario  must necessarily (for all reasonable parameters) have luminosities  as low as in Marley et al. (\cite{marleyfortney2007}) (see also the recent calculations of Spiegel \& Burrows \cite{spiegelburrows2011}). Furthermore,  the results in the next two sections show that the formation of giant planets via core accretion, but with ``hot accretion'' leads to high initial luminosities which very quickly converge on the luminosity of classical ``hot start'' calculations (Burrows et al. \cite{burrowsmarley1997};  Baraffe et al. \cite{baraffechabrier2003}). We thus see that depending on currently ill constrained physical mechanisms occurring during the formation phase (in particular the nature of the accretion shock),  the luminosity of planets formed by core accretion can probably vary significantly.} 

\subsubsection{Gravitational instability}
{For the second formation mechanism, the  gravitational instability  (GI) model, the question about the associated luminosity might also be relatively complex as indicated by the explorative discussion in this section. More solid results are expected once  multidimensional, radiation-hydrodynamic calculations following the entire formation by gravitational instability from the first instability in the disk to a compact planet at the final mass will become feasible.}

{In a qualitative way, we could expected that the post-formation luminosity of planets formed by GI depends on the specific way these objects grow in mass: Hydrodynamic simulations show that the initial mass of the clump which becomes self-gravitationally unstable is typically 3-20 $\mj$ (e.g. Forgan \& Rice \cite{forganrice2011}; Helled \& Bodenheimer \cite{helledbodenheimer2011}; Zhu et al. \cite{zhuhartmann2012}). These masses roughly correspond to the ones expected analytically from the mass contained in the disk patch with a length scale similar to the most unstable mode. The objects that form at this stage are self-gravitating, relatively cold, extended objects. Their radius is comparable to the Hill sphere radius (e.g. Zhu et al. \cite{zhuhartmann2012}). For example, a 10 $\mj$ clump at 50 AU from a 1 $\msun$ star which may represent a typical outcome (e.g. Janson et al. \cite{jansonbonavita2011}) has a Hill sphere radius of about 7.4 AU. These clumps consist of molecular hydrogen and, once formed, contract slowly (e.g. Helled \& Bodenheimer \cite{helledbodenheimer2011}). In their nature they  correspond to the first Larson core of star formation.}

{The mass contained in such an object has not gone through an accretion shock, since they form from the spontaneous self-gravitational binding of a  extended part of the disk, typically within  an over-dense spiral arm (e.g. Boley et al. \cite{boleyhayfield2010}). We therefore expect that the specific entropy of the gas in the clump should be high. This is confirmed by numerical simulations: T. Hayfield (personal communication, 2012, see also Galvagni et al. \cite{galvagnihayfield2012}) finds that the specific entropy in a clump taken from the radiation-hydrodynamic simulations of Boley et al. (\cite{boleyhayfield2010}) is about 15 $k_{\rm B}$/baryon at the moment when the clump should undergo the second collapse (cf. below). If this is representative of the entropy of the final object, this corresponds to a "hottest start" (Spiegel \& Burrows \cite{spiegelburrows2011}). This allows to make a first supposition: If the final mass of a GI planet is the same (or similar) as the mass of the initially unstable clump (an approximation which is regularly made, e.g. Janson et al. \cite{jansonbonavita2011}), then we may expect that these objects have a ``hot start", simply because there is no accretion shock involved during their formation.}

{However, since the clump necessarily forms in a massive disk, it appears more likely that the clump continues to accrete mass (in principle, gap formation can stop accretion, e.g. Zhu et al. \cite{zhuhartmann2012}). In this case, we have to distinguish two sub-scenarios. As the first core contracts quasi-statically, its interior heats up, until a central temperature of about 2000 K is reached. Then, molecular hydrogen dissociates, and a dynamical collapse of the entire clump occurs (e.g. Bodenheimer et al. \cite{bodenheimergrossman1980}). This correspond to the formation of the second core in star formation. The timescale until this second collapse occurs is estimated to be roughly $10^{3}$ to $10^{4}$ years, with more massive clumps collapsing earlier (Helled \& Bodenheimer \cite{helledbodenheimer2011}).} 

{In the first sub-scenario the clump grows to its final mass predominately before it collapses. Accretion in this phase is efficient, since the cross section is approximately given by the large Hill sphere radius. This results in high to very high accretion rates of  $10^{-7}$ to $10^{-4} \msun$/yr  (Boley et al. \cite{boleyhayfield2010}; Zhu et al. \cite{zhuhartmann2012}). The object therefore may grow quickly out of the planetary mass regime (Kratter et al. \cite{krattermurrayclay2010}). Independently of this issue, for the ``hot'' vs. ``cold accretion'' question, it is relevant in which way the gas is incorporated into the clump, in particular if it is accreted via a supercritical shock. The escape speed from the Hill sphere of a 10 $\mj$ object at 50 AU is about 1.5 km/s. For comparison, in the calculation of the accretion onto the first (stellar) core in Commer\c{c}on et al. (\cite{commerconaudit2011}), the shock velocity is about 2.4 km/s, which is comparable. Because we are in a shearing disk, both Boley et al. (\cite{boleyhayfield2010}) and Zhu et al. (\cite{zhuhartmann2012}) however find that the relevant velocity with which they can reproduce the numerically obtained accretion rates is not the free fall velocity, but the Keplerian shear velocity across the Hill sphere. The  velocity with which the newly accreted matter joins the clump is thus of order $\Omega R_{\rm H}$ where $\Omega$ is the Keplerian frequency. In the example, this corresponds to about 0.6 km/s. Whether or not this is sufficient to produce a supercritical shock must be evaluated for realistic velocities, pre-shock densities and temperatures. If it is not, as we may suspect due to the rather low velocity, then the material accreted in this phase should be added at a high specific entropy. This leads to the second supposition: If a GI planet accretes most of its final mass before it undergoes the second collapse, and the accretion onto the AU sized clump is relatively gentle, we may again expect a  ``hot start".  }

{In the second sub-scenario, the clump accretes most of its final mass after it has undergone the second collapse which leads to an object with a size of a few Jovian radii (Helled \& Bodenheimer \cite{helledbodenheimer2011}). Now, the accreted gas hits the protoplanet at a much higher velocity of several 10 km/s as in the runaway gas accretion phase in the core accretion scenario. This leads to the third supposition: The larger the fraction of the final mass in a GI planet accreted after the clump has undergone the second collapse, the ``colder" it should be, provided that the accretion shock in the post-collapse phase is supercritical. The effect that the higher the fraction of the final mass processed through the shock, the lower the initial luminosity, is seen in the simulations of Marley et al. (\cite{marleyfortney2007}) where it causes more massive planets to be less luminous than their lighter counterparts (``luminosity inversion''). }

{This  discussion (see also the related discussion in Spiegel \& Burrows \cite{spiegelburrows2011}) shows that with the currently limited knowledge both about the specific way matter is accreted, and the involved thermodynamics in general, it is difficult to firmly associate a ``hot/cold start'' with a specific formation mechanism. Some of the general trends for the impact of formation on the initial luminosity of core accretion planets will be presented in a dedicated paper (Mordasini et al. in prep.). }

\subsection{``Hot start'' and ``cold start'' models} 
\begin{figure*}
\begin{minipage}{0.5\textwidth}
	      \centering
       \includegraphics[width=1\textwidth]{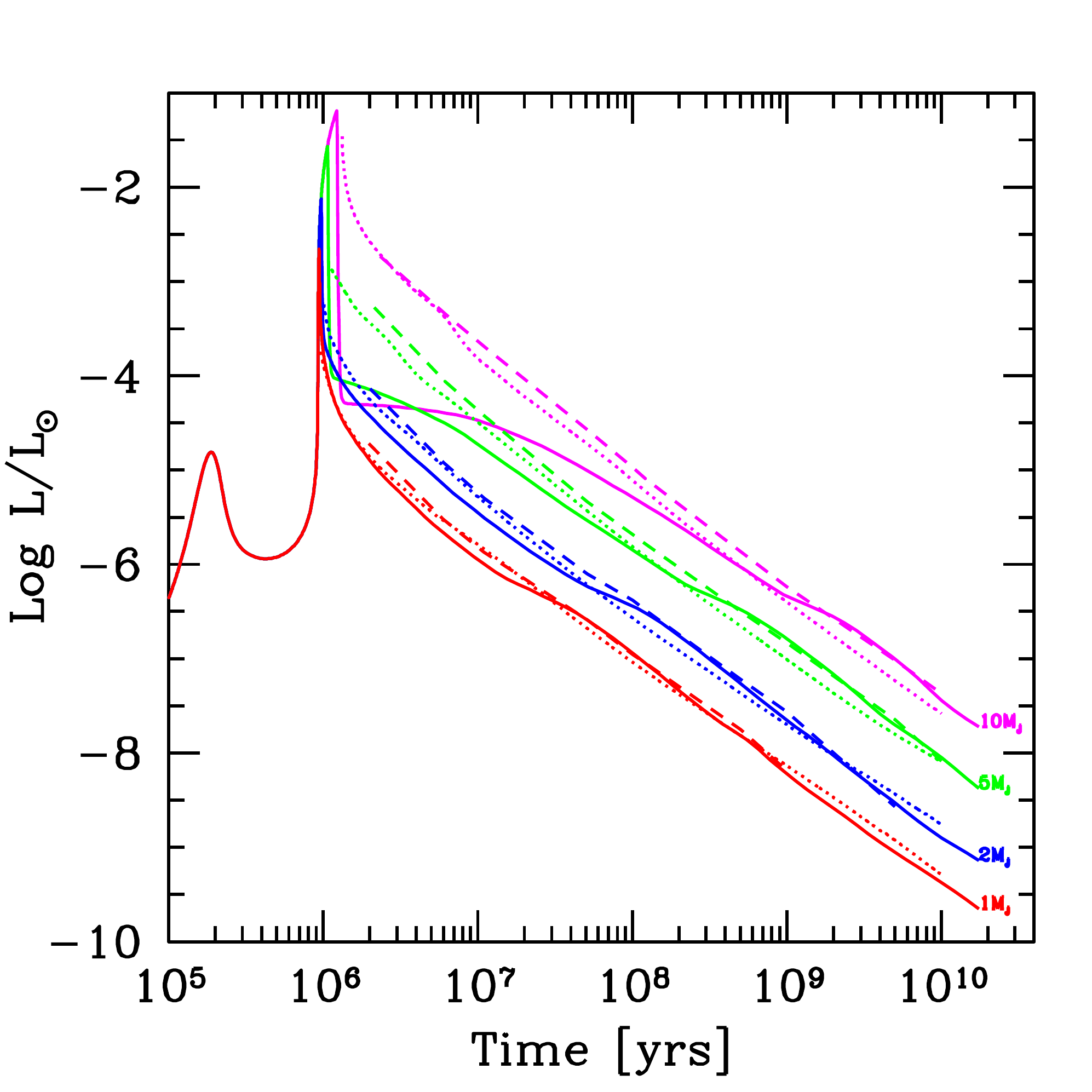}
                     
     \end{minipage}
     \begin{minipage}{0.5\textwidth}
      \centering
       \includegraphics[width=1\textwidth]{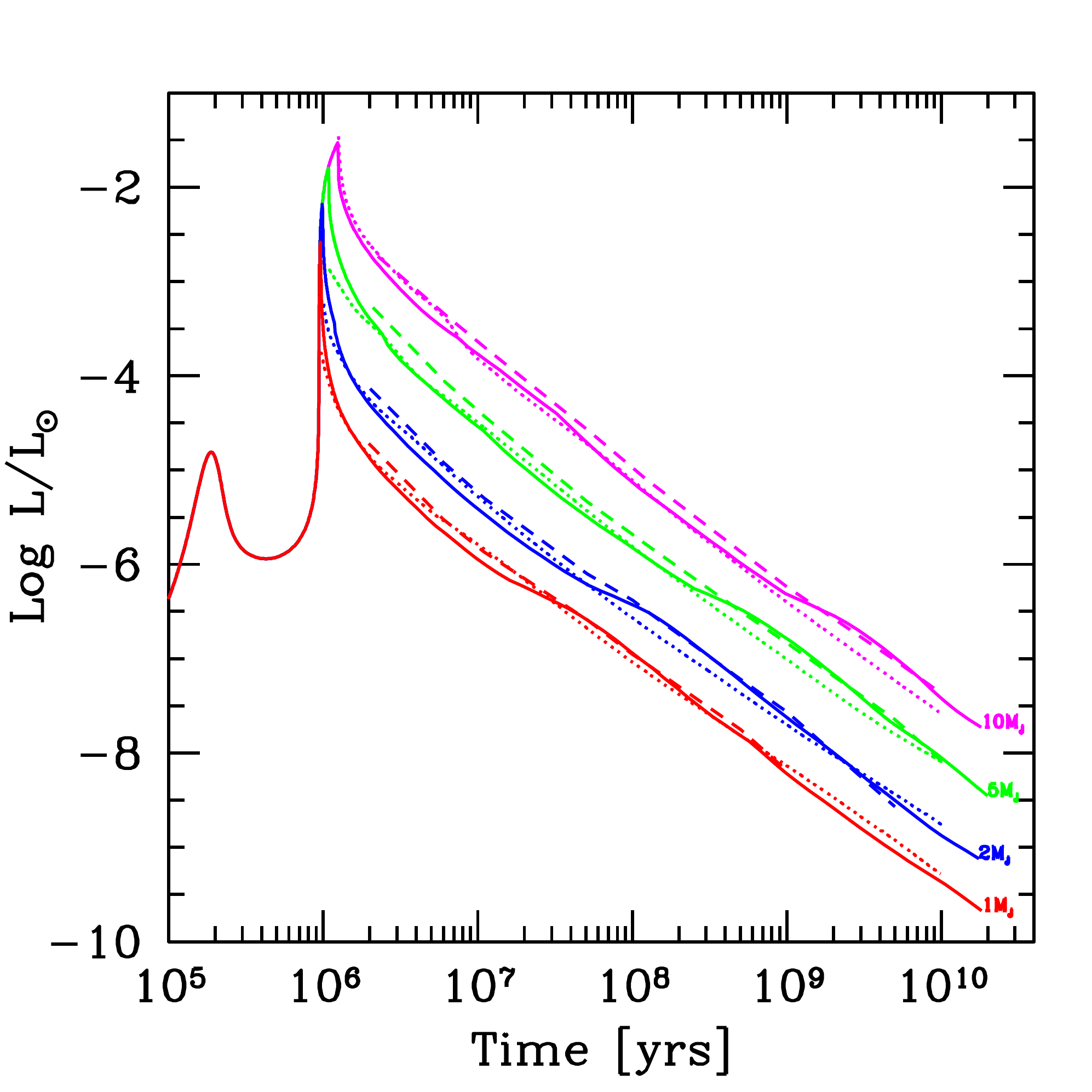}
     \end{minipage}
             \caption{Luminosity as a function of time for planets with a mass of 1, 2, 5 and 10 $\mj$, as labelled in the plot. The left panel shows the case of ``cold start'' boundary conditions where all accretional energy is radiated away at the shock, while the right panels shows ``hot start (accreting)'' models where no radiative losses occur. The dotted lines show for comparison the results of Burrows et al. (\cite{burrowsmarley1997}), while the dashed lines show Baraffe et al. (\cite{baraffechabrier2003}). Both these models use the classical ``hot start'' scenario. The ``cold start'' 1 $\mj$ simulation is the same as shown in Fig. \ref{fig:ltlong}. }
             \label{fig:tlumich} 
\end{figure*}

In the present paper, we will only compare our results with already published studies, similar as for the radii in the previous section. Figure \ref{fig:tlumich} shows the (total) luminosity (including the accretional luminosity)  as a function of time for planets of 1, 2, 5 and 10 $\mj$.

The left panel shows the result using ``cold start'' boundary conditions, i.e. it is assumed that all accretional energy is radiated away at the shock, while no losses are assumed to occur in the ``hot start (accreting)'' case (Eq. \ref{eq:hotcoldl}). The cold start, $M=1 \mj$ simulation is the same one as discussed in detail in Sect. \ref{sect:examplesinsitu}, while the other masses correspond to the simulations shown in Fig. \ref{fig:tmdiffm}.  We plot in the figures also the $L(t)$ found by  Burrows et al. (\cite{burrowsmarley1997}) and Baraffe et al. (\cite{baraffechabrier2003}){, for comparison\footnote{The data plotted is from \href{http://www.astro.princeton.edu/~burrows/}{http://www.astro.princeton.edu/$\sim$burrows/} and \href{http://perso.ens-lyon.fr/isabelle.baraffe/}{http://perso.ens-lyon.fr/isabelle.baraffe/}.}.} Both these models are classical ``hot start'' simulations. We associate the ``$t=0$'' moment of these models with the moment when in our simulations, the planets  reach their final mass.  For the ``cold start'' simulations, this moment  corresponds to the  sharp drop of $L$ particularly well visible for the 10 and 5 $\mj$ cases, when $L_{\rm acc}$ vanishes.

\subsubsection{``Hot start'': comparison with Burrows et al. (\cite{burrowsmarley1997}) and Baraffe et al. (\cite{baraffechabrier2003})}\label{sect:hotstartcompbb}
Focussing first on the right panel with the ``hot start (accreting)'' models, we see a good agreement between our model and the two other ones. The differences between our (simpler, grey atmosphere) model and the two more complex models is of the same order as the differences mutually between Burrows et al. (\cite{burrowsmarley1997}) and Baraffe et al. (\cite{baraffechabrier2003}). This is in agreement with  Bodenheimer et al. (\cite{bodenheimerhubickyj2000}) who also find very good agreement of their grey atmosphere models and Burrows et al. (\cite{burrowsmarley1997}). It is however clear that the precise shape of $L(t)$ (there is for example a small bump in our cooling curves when  $\log L/L_{\odot}\approx -6.25)$ depends directly on the opacities, so that we expect that our predictions are somewhat less accurate during the long-term evolution. 

{One} sees that the assumption of no radiative losses at the shock, but still a gradual building up of the planet (as in our simulations) leads to very similar results as the classical ``hot start'' scenario where one starts with a fully formed planet. The physical reason for the similarity is that in both cases, no entropy sink exists,  {and that the resulting short Kelvin-Helmholtz timescales make that the initial conditions are quickly ``forgotten'' in a rapid convergent evolution, as discussed at the end of this section}. {It means that core accretion with a radiatively completely inefficient shock leads to very similar planetary luminosities at young ages as in classical ``hot start'' simulations.}

This result certainly underlines the importance of a detailed description of the shock structure, as already pointed out by Marley et al. (\cite{marleyfortney2007}), because we see that the shock structure has an influence which is as important as the formation mechanism. Progress on the shock structure is an important task for future studies (see Commer\c{c}ont et al. \cite{commerconaudit2011} for a recent study of the stellar case).  

{Note that the assumption of $dl/dr=0$ in the envelope for the ``hot start'' scenario could lead to significant departures from the actual interior structure and the associated luminosity of the planets  during  the gas runaway accretion phase. This is due to the fact that we cannot catch  the  effects of a possible deep radiative zone, as discussed in Sect. \ref{sect:issuehotdldr0}.  This would also affect the luminosities at the beginning of the evolutionary phase i.e. just after the end of formation.} 

{The comparison of the luminosity at this moment as found in our simulations, and as assumed in Burrows et al. (\cite{burrowsmarley1997}) and Baraffe et al. (\cite{baraffechabrier2003}) indicates that we have luminosities which are similar (for the $10\mj$ case) or higher (for lower masses). A higher luminosity means that the Kelvin-Helmholtz timescale is very short, so that the models converge on a timescale of only $10^{5}-10^{6}$ yrs. Such a rapid convergence of ``hot start'' and ``hotter start'' models has been found by Baraffe et al. (\cite{baraffechabrier2002}, \cite{baraffechabrier2009}) (see also Marley et al. \cite{marleyfortney2007} for a discussion). The effect of a variable luminosity on ``hot start'' objects, calculated as described in Sect. \ref{sect:testsimplidldr0}, will be studied in future work. This should allow to better describe the properties of very young ``hot start'' objects.}

\subsubsection{``Cold start'': comparison with Marley et al. (\cite{marleyfortney2007})}\label{sect:coldstartcompmarley}
Comparing the left and the right panel makes it immediately clear that we recover the important findings of Marley et al. ({\cite{marleyfortney2007}) (which were  already visible in the simulations of Bodenheimer et al. \cite{bodenheimerhubickyj2000}), that the luminosity of young planets formed according to the ``cold start'' assumption can be substantially lower, at least for high-mass planets. 

Focussing  on the left panel with the ``cold start'' simulations we see that we recover the most important  effects found by Marley et al. ({\cite{marleyfortney2007}): we also see a ``luminosity inversion'' which means that the post-accretional luminosity (i.e. the luminosity shortly after the final mass of the planet is reached) is the highest for the lowest mass planet. The picture is somewhat complicated by the fact that the low-mass planets start with the highest $L$ but also cool the quickest, as their KH-timescale $G M^{2}/ (R L)$ is the shortest. The reason for the ``luminosity inversion'' is that the higher the total mass, the higher the fraction of gas in the envelope that has been process through the entropy-reducing shock.  The total mass of the planets at the moment when they start to collapse is about 87 $\mearth$, with a core mass of 22 $\mearth$ and an envelope of 65 $\mearth$ (cf. Table \ref{tab:Jx}).  This is equal for all planets (their evolution is identical up to the point  where  for the lower mass planets, accretion is shut off). This means that for a $1 \mj$ planet, about 20\% of its final envelope mass is accreted in the attached phase, and thus without going through the entropy-reducing shock. The ratio is proportionally smaller for higher mass planets, leading to the inversion.

Another identical result is that the higher the mass, the longer it takes until ``hot start'' and ``cold start'' luminosities become equal, a result that can be understood from the KH-timescales. For the 10 $\mj$ planet, it takes several 100 Myrs, similar to Marley et al. ({\cite{marleyfortney2007}).

Quantitatively, there are however also some differences: we note that we find post-accretional luminosities for the massive (5 and 10 $\mj$) planets that are of order $\log L/L_{\odot}\approx-4.1$ to $-4.3$, while Marley et al. ({\cite{marleyfortney2007}) rather find $\log L/L_{\odot}\approx -5.7$, i.e. quite fainter. The reason for this difference will be investigated in detail in Mordasini et al. (in prep.).

\section{Summary}
We have extended our formation model to a self-consistently coupled formation and evolution model for planets with a primordial H$_{2}$/He envelope. In this paper we have described the modification we made to the computational module that calculates the internal structure in order to make this extension possible. We then compared the results concerning planetary evolution with existing work. We found good agreement with more complex models. In a companion paper we describe further extensions and improvements relevant during the evolution of the planets, like an internal structure model for the solid core assuming a differentiated planet which gives realistic radii also for planets without significant atmospheres, or the radiogenic luminosity of the planet's mantle. {In the companion paper we also  discuss extensively our results on  planetary radii as obtained in population synthesis calculations.}

\subsection{Jupiter: coupled in situ formation and evolution}
We have simulated the combined formation and evolution of Jupiter in the framework of classical core accretion models without migration and disk evolution (Pollack et al. \cite{pollackhubickyj1996}). We have shown that the upgraded model with the new simple and rapid method to calculate evolutionary sequences leads to results which are in very good agreement with those of several other models (Lissauer et al. \cite{lissauerhubickyj2009}; Burrows et al. \cite{burrowsmarley1997}; Baraffe et al. \cite{baraffechabrier2003}). We have found that the nominal model for Jupiter leads to the formation of a planet which has properties at 4.6 Gyrs which are in excellent agreement with observed values in terms of internal composition, radius, luminosity and surface pressure-temperature profile.

\subsection{Planetary radii}
We have studied the  mass-radius diagram of giant planets and the influence of the core mass on the total radius.  We have found that the upgraded models yields planetary radii of giant planets which are in very good agreement with more sophisticated models (Fortney et al. \cite{fortneymarley2007};  Baraffe et al. \cite{baraffechabrier2008}). We must assume that our cooling curves are probably less accurate for low-mass, core-dominated planets, possibly with heavily  enriched envelopes. Still we show in the companion Paper II that the models yields also for a completely different type of planet, namely a close-in, $4.2\mearth$ super-Earth planet with a tenuous $\sim1\%$ H$_{2}$/He envelope a radius which agrees with the more detailed simulations of Rogers et al. (\cite{rogersbodenheimer2011}) to better than 10\%, corresponding to about 0.2 $\rearth$. We thus can calculate planetary radii for a very wide range of planets, which is our goal for the planet population synthesis calculations for which the model is eventually intended (see Paper II).  

\subsection{Planetary luminosities}
We have used our upgraded model to study the luminosity of giant planets of different masses as a function of time. We have made calculations both for the ``cold start'' and the ``hot start (accreting)'' scenario. For the  ``hot start (accreting)'' models we have found cooling curves which are very similar as in Burrows et al. (\cite{burrowsmarley1997}) and Baraffe et al. (\cite{baraffechabrier2003}) despite the fact that we use simple gray boundary conditions.  In the ``cold start'' calculations we have recovered the result from Marley et al. (\cite{marleyfortney2007}) that the luminosities of massive young planets are much smaller, at least for the chosen parameters. In a dedicated work (Mordasini et al. in prep.) we revisit the luminosity of young Jupiters using a large suite of combined formation and evolution calculations.

\section{Conclusion}
In earlier versions of the model, we calculated the internal structure of the gaseous envelope only during the attached, pre-gas runaway accretion phase. This is sufficient if one is only interested in the final mass of the planets, and thus sufficient for comparisons with planets found by the radial velocity method. Now we calculate the structure also during the gas runaway accretion phase (which is also the phase when the planet's radius collapses) and during the evolutionary phase at constant mass over Gigayear  timescales. With that we now know all major quantities  (mass, semimajor axis, radius, luminosity, composition) characterizing the planets during their entire formation and evolution. This allows to  compare our population synthesis models directly and consistently with results coming from all major observational techniques used to detect and characterize extrasolar planets {(see Paper II for a comparison of the synthetic and actual mass-radius relationship,  the predicted distribution of radii, and the comparison with \textit{Kepler})}. 

Extrasolar planet research has entered an era in which massive amounts of observational data regarding very different sub-populations of planets are brought to us from different techniques. They should all be explained consistently by planet formation and evolution theory, but it is a difficult task to unite the different elements and observational constraints into one consistent global picture. 

We think that  a  fruitful approach in this situation  is to work with theoretical models able to make testable predictions in a consistent way for all important observational techniques. With the work presented in this paper we make a development in this direction, allowing us to test and improve theoretical formation models using the wealth of data coming from observations.

\acknowledgements{We thank Christopher Broeg for dedicated comparison calculations. We thank J\'er\'emy Leconte for answering many questions related to giant planet evolution. We thank Willy Benz, Gilles Chabrier, Isabelle Baraffe, Chris Ormel, Andrea Fortier and Tristen Hayfield for helpful discussions. {We thank an anonymous referee for questions which have helped to illuminate the impact of our assumptions.} Christoph Mordasini acknowledges the support as an Alexander von Humboldt fellow. This work was supported in part by the Swiss National Science Foundation and the European Research Council under grant 239605.  Computations were made on the BATCHELOR cluster at MPIA.}

\appendix
\section{Comparison calculation with Pollack et al. (\cite{pollackhubickyj1996})}\label{comparisonp96}
In this appendix, we present a detailed comparison calculation with the classical simulation of Pollack et al. (\cite{pollackhubickyj1996}). In particular, we simulate the cases J1, J1a, J1b and J1c. Simulation J1 is the nominal case, while for the other cases, solid accretion gets (arbitrarily) shut off at specified times (see Figure \ref{fig:p96comp}). The consequence of this is a reduced core luminosity, which in turn leads to a faster gas accretion due to the reduced thermal support of the envelope, an effect described by Pollack et al. (\cite{pollackhubickyj1996}). 

This effect is relevant in our population synthesis calculations in two situations: during planet migration, and/or the concurrent formation of several planets. The reason  for the first situation is that with the updated (non-isothermal type I) migration model, protoplanets are often seen to first migrate outwards, and then back in. On their way back in, they move into the region of the planetesimal disk which they have cleared from planetesimals while migrating outwards (see Mordasini et al. \cite{mordasinidittkrist2011}). In the second situation, one planet can migrate into a region which has previously been cleared by another growing planet. In both situations, a faster gas accretion will result. From this we see that migration and accretion can interact. 

With this appendix, we want to check whether our simplified luminosity calculation reproduces the effect found by P96, and in general compare quantitatively our results with the published ones.

The J1 initial conditions in particular mean that the initial planetesimal surface density $\Sigma_{\rm p}$ is 10 g/cm$^{2}$ as in Section \ref{sect:examplesinsitu}.  The grain opacities are now however at 100 \% of the interstellar value ($\fopa=1$). The density of the core is constant at 3.2 g/cm$^{3}$ as in Pollack et al. (\cite{pollackhubickyj1996}), and also the nebular boundary conditions are the same as in this work. For the J1a, J1b and J1c cases, we switch off planetesimal accretion at the same moments as in Pollack et al. (\cite{pollackhubickyj1996}). The parameters used for the simulations are given in Table \ref{tab:j1comp}. It is clear that the two simulation still cannot yield exactly identical results, as they differ in some other aspects, like a different equation of state, or a different model for the planetesimal-envelope interaction. 

\begin{table}
\caption{Settings for the Pollack et al. (\cite{pollackhubickyj1996}) comparison calculations.}\label{tab:j1comp}
\begin{center}
\begin{tabular}{lc}
\hline\hline
case & J1 comparison\\ \hline
a [AU] & 5.2 \\
 $\sigmas0$ [g/cm$^{2}$]                         & 10       \\                                              
$\dot{M}_{\rm XY,max}$ [$\mearth$/yr]& 0.01 \\
$T_{\rm neb}$ [K]           &150\\
$P_{\rm neb}$ [dyn/cm$^{2}$]           &   0.275      \\ 
Dust-to-gas ratio  &1/70  \\\hline
Initial embryo mass [$\mearth$] & 0.1\\
Migration & not included\\
Disk evolution & not included\\
Planetesimal ejection & not included\\
Core density & constant, 3.2 g/cm$^{3}$\\
Simulation duration & $10^{10}$ yrs\\
Grain opacity red. factor $\fopa$ & 1 \\ \hline
\end{tabular}
\end{center}
\end{table}

In Figure \ref{fig:p96comp} we show the core mass, envelope mass, and total mass for the four models, and in Table \ref{tab:Jx} we give the values of the most important quantities at specific moments in time. We first focus on the nominal J1 case.  The crossover point, i.e. the moment when the core and the envelope mass are identical and equal to $M_{\rm cr}$ is found to occur at $t_{\rm cr}=7.49$ Myrs at $M_{\rm cr}=16.62\mearth$. This is very close to   Pollack et al. (\cite{pollackhubickyj1996}) with $t_{\rm cr}$=$7.58$ Myrs and $M_{\rm cr}$=$16.17\mearth$. Note however that the very good agreement in $t_{\rm cr}$ is partially a chance result, because it is well known that changing for example just the opacity tables or the EOS can lead to variations of several million years (Hubickyj et al. \cite{hubickyjbodenheimer2005}). These  authors found for identical conditions, but an updated version of the  model used in Pollack et al. (\cite{pollackhubickyj1996}), a $t_{\rm cr}$=6.07 Myrs and $M_{\rm cr}$=$16.16\mearth$. At crossover, the core and gas accretion rate are $2.37\times 10^{-6}$ and $1.19\times10^{-5}\mearth$/yr, which are both in very good agreement with Pollack et al. (\cite{pollackhubickyj1996}) .

\begin{figure}
\begin{center}
\includegraphics[width=\columnwidth]{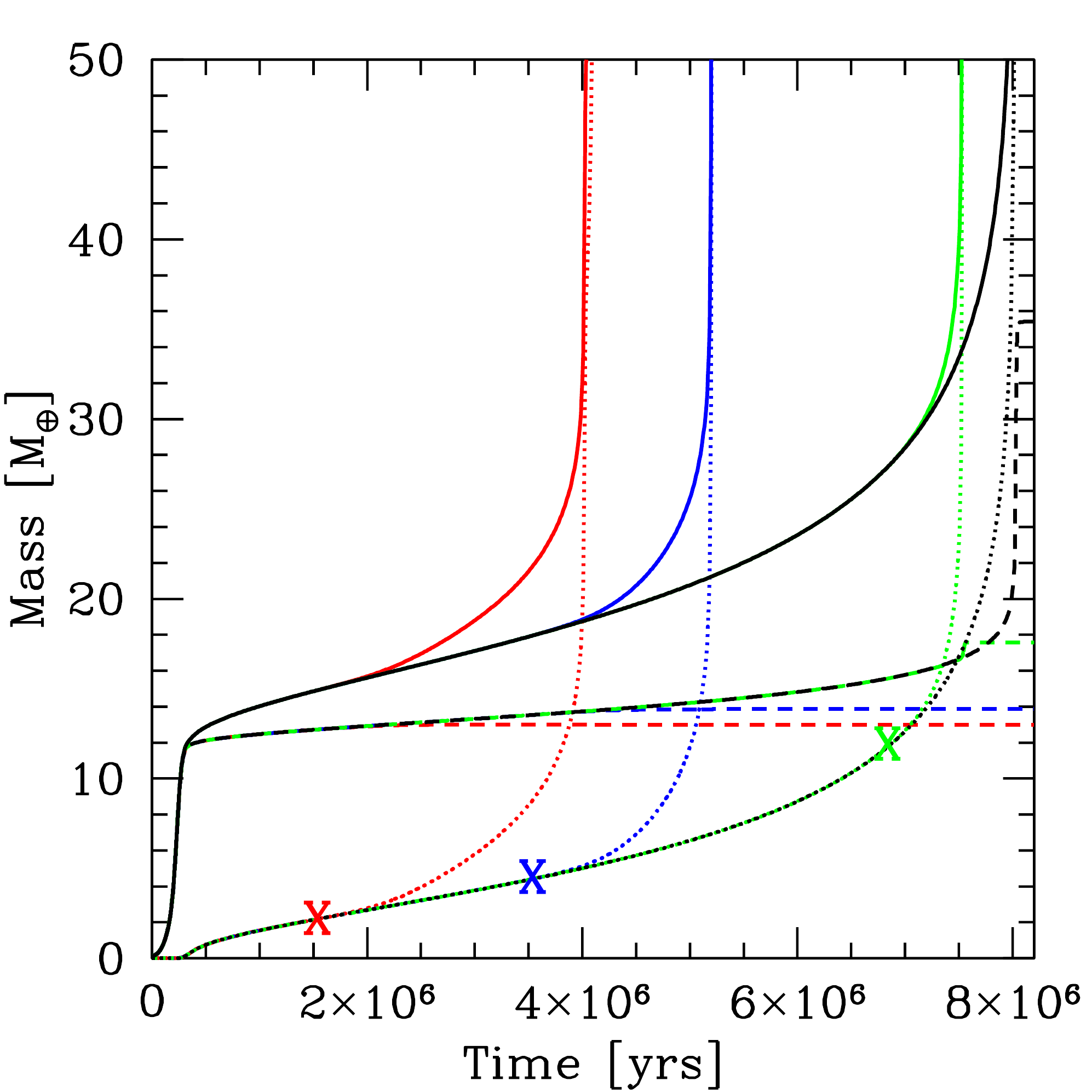}
\caption{Total mass (solid line), envelope mass (dotted line) and core mass (dashed line) for the cases J1 (black), J1a (red), J1b (blue) and J1c (green).  The X-symbols indicate the moment in time when $\mdotz$ is artificially ramped down for the latter three simulations. } 
\label{fig:p96comp}
\end{center}
\end{figure}

The limiting gas accretion rate of 0.01 $\mearth$/yr is reached at 8.02 Myrs, about half a million year after crossover. The core mass is then about 22.90 $\mearth$ and the envelope mass 75.8 $\mearth$. In Pollack et al. (\cite{pollackhubickyj1996}), a similar, but somewhat smaller gas accretion rate is reached 0.42 Myrs after crossover, at $\mz=21.5\mearth$ and $\mxy=64.4\mearth$.  In contrast to Pollack et al. (\cite{pollackhubickyj1996}), we have taken the calculation also through the detached end evolutionary phase. The second luminosity peak occurs shortly after the limiting gas accretion is reached, at 8.036 Myrs, when the planet has already contracted to a radius of 4.49 $\rj$, and grown to a mass of 267 $\mearth$. The luminosity is then $\log(L/L_{\odot})=-2.77$. Lissauer et al. (\cite{lissauerhubickyj2009}) find peak luminosities for planets accreting with the same  limiting gas accretion rate of  $\log(L/L_{\odot})\approx-2.4$ to $-2.3$. 

The case J1a (red lines) is identical to case J1, except that at $t=1.5$ Myrs, the solid accretion rate is artificially ramped down to zero on a short timescale as in Pollack et al. (\cite{pollackhubickyj1996}). This results in a faster growth of the envelope, i.e. an increase of the gas accretion rate. Our model thus reproduces this important effect. In this simulation, crossover occurs already at $t_{\rm cr}$=3.38 Myrs and $M_{\rm cr}$=$12.98\mearth$. In Pollack et al. (\cite{pollackhubickyj1996}), the values are 3.32 Myrs, and  12.24 $\mearth$ which shows that also quantitatively, there is very good agreement.  The results for the other two cases, J1b and J1c, are comparable, even though the differences in timescales are somewhat larger. 

In summary we see that there is also quantitatively a very good agreement between the Pollack et al. (\cite{pollackhubickyj1996}) simulations and the ones presented here.

\begin{table}\scriptsize 
\caption{Models for the in situ formation of Jupiter. J1 and J1a are the comparison calculations with Pollack et al. (\cite{pollackhubickyj1996}). The third column is our own nominal model for Jupiter's in situ formation presented in Section \ref{sect:examplesinsitu}. It should be more realistic because it includes a reduced grain opacity, a variable core density and the ejection of planetesimals. }\label{tab:Jx}
\begin{center}
\begin{minipage}[l]{8.9cm}
\begin{tabular*}{8.9cm}{llrr|r}
\hline
Phase&  & J1 & J1a & nominal\\
\hline
First luminosity peak & Time 	 & 0.242 & 0.242  & 0.188  \\
        & $M$ & 8.13& 8.13&8.20 \\
                   & $\mz$ &8.12&8.12 & 8.17\\
                             & $\mxy$&0.01 &0.01 & 0.03 \\
                             & $\mdotz$ &1.00$\times$$10^{-4}$ & 1.00$\times$$10^{-4}$& 1.40$\times$$10^{-4}$ \\
                                   
                                     & $R$ & 25.8&25.8 & 26.9\\

    				                 & $\log L/L_{\odot}$ & -4.99 & -4.99 & -4.81 \\ \hline
 							
 Cross over point                  & Time & 7.49 & 3.38  & 0.817  \\
                   & $M$ & 33.23& 25.96&32.53 \\
                                 & $M_{\rm cr}$ &16.62 & 12.98& 16.27\\
                                    & $\mdotz$ &2.37$\times$$10^{-6}$ & 0 & 1.70$\times$$10^{-5}$ \\
                                      & $\mdotxy$ &1.19$\times$$10^{-5}$& 2.14$\times$$10^{-5}$ &8.33$\times$$10^{-5}$ \\
                                     & $R$ & 61.2& 53.5  & 60.4\\
	 	                   & $\log L/L_{\odot}$ & -6.31&-6.84 & -5.46\\\hline
					             
Onset of   limited                        & Time & 8.02 & 4.125  & 0.923  \\
  gas   accretion                    & $M$& 98.7& 107.5& 86.7\\
                                         & $\mz$ &22.9 & 13.0 &  21.8\\
                                       & $\mxy$&75.8 &94.5  & 64.9 \\
                                               & $\mdotz$ & 5.94$\times$$10^{-4}$ & 0 &  6.13$\times$$10^{-4}$\\
                                                          & $R$ & 104.0 & 108.2 & 98.1 \\
 	                                           & $\log L/L_{\odot}$ & -4.03 & -4.10 & -4.03  \\ \hline 
							                     
Second  luminosity      peak                               & Time & 8.036 & 4.141 &  0.940  \\
                                    & $M$ & 266.9  & 266.0 & 264.4 \\
                                             & $\mz$ &  32.7 & 13.0& 30.9 \\
                                                    & $\mxy$& 234.3 & 252.9   &233.5 \\
                                                     & $\mdotz$ &4.84$\times$$10^{-4}$   &  0  & 3.60$\times$$10^{-4}$\\
                                                      & $R$ & 4.49  & 2.32 &  3.77\\
                                            	   & $\log L/L_{\odot}$ &-2.77  & -2.54  &-2.65 \\ 
							 	                  
                                                       & $T_{\rm cent}$ & 5.25$\times$$10^{4}$  & 5.01$\times$$10^{4}$ & 6.08$\times$$10^{4}$ \\ \hline

Present day   (4.6 Gyr)              
                       & $M$& 318.7& 317.8   & 316.6\\
                                    & $\mz$ & 35.4& 13.0 &32.9 \\
                                 & $\mxy$& 283.3 & 304.8 &  283.7\\
                                      & $R$ & 1.14& 1.15  & 0.99\\
                                     & $\rho_{\textrm core}$ & 3.2 &  3.2& 14.31\\
 			         & $\log L/L_{\odot}$ &  -8.83 & -8.79  & -9.01 \\ 
			                  & $L/\lj$ &  1.73  &  1.88 & 1.13 \\ 
			                 & $T$ & 128.4  & 129.2 & 126.3\\ 
			                & $T_{\rm cent}$ & 2.16$\times$$10^{4}$  & 2.28$\times$$10^{4}$  & 1.76$\times$$10^{4}$\\ \hline

\end{tabular*}
\end{minipage}

\begin{minipage}[l]{9cm}
The following units are used: Time in Myrs; Masses in $\mearth$; Accretion rates in $\mearth/yr$; Radii in Jovian equatorial radii  equal $7.15\times10^{9}$ cm; Core densities in g/cm$^{3}$; Temperatures in K; $\lj$ is the present day Jovian intrinsic luminosity $\lj=8.7\times10^{-10}\lsun$.
\end{minipage}
\end{center}
\end{table}

\end{document}